\documentclass[twocolumn, trackchanges, floatfix]{aastex62}
\hypersetup{linkcolor=black,citecolor=blue,filecolor=cyan,urlcolor=magenta}
\usepackage{amsmath}

\graphicspath{{./}{figures/}}

\shorttitle{AGN driven outflows in dwarf galaxies}
\shortauthors{Manzano-King et al.}

\begin{document}

\title{AGN-Driven Outflows in Dwarf Galaxies}

\correspondingauthor{cmk}
\email{cking012@ucr.edu}

\author[0000-0002-5253-9433]{Christina M. Manzano-King}
\affil{Department of Physics and Astronomy\\
University of California Riverside\\
900 University Ave. CA 92521, United States}

\author[0000-0003-4693-6157]{Gabriela Canalizo}
\affil{Department of Physics and Astronomy\\
University of California Riverside\\
900 University Ave. CA 92521, United States}

\author[0000-0002-3790-720X]{Laura V. Sales}
\affil{Department of Physics and Astronomy\\
University of California Riverside\\
900 University Ave. CA 92521, United States}

\begin{abstract}

We present spatially resolved kinematic measurements of AGN-driven outflows in dwarf galaxies in the stellar mass range $\sim6\times10^8 - 9\times10^9 M_\odot$, selected from SDSS DR7,8 and followed up with Keck/LRIS spectroscopy.  We find spatially extended ($\sim 1$ half light radius), high velocity ionized gas outflows ($W_{80}$ up to $\sim2000\,\rm{km\,s}^{-1}$) in 13/50 dwarf galaxies with and without AGN.
Outflow velocities in all 13 galaxies exceed the escape velocities of their halos.  Nine of these 13 galaxies are classified as AGN according to their narrow line flux ratios. Of these, six have outflow components with emission line ratios consistent with AGN ionization.  Although black holes have been known to populate the centers of at least a few dwarf galaxies, and indirect evidence of AGN quenching of star formation in dwarfs has begun to surface, our measurements constitute the first direct detection and measurement of AGN impact on the large scale kinematics and gas content in dwarf galaxies.  Furthermore, we find evidence suggestive of ongoing star formation suppression, possibly regulated by the AGN.  Galaxy formation models must therefore be able to account not only for the formation and growth of black holes at the centers of dwarf galaxies, but should also be revised to include AGN as important –-and perhaps dominant-– sources of feedback in low mass galaxies.

\end{abstract}



\section{Introduction} \label{sec:intro}

Well established correlations between black hole (BH) masses and global galactic properties imply a scenario where the growth and evolution of a BH and its host galaxy are connected and regulated by feedback from the active galactic nucleus (AGN) (e.g., \cite{Kormendy_Ho_2013}).  In massive galaxies, the presence of AGNs has been shown to be linked to the development of powerful gas outflows (e.g., \citet{Harrison2014, McElroy2015, Cheung2016, Rupke2017}), which are a vital ingredient of galaxy formation models within the cosmological framework of $\Lambda$ Cold Dark Matter ($\Lambda$CDM), as they regulate and suppress star formation \citep{Vogelsberger2014,Schaye2015,Dubois2016,Pillepich2018} and bring the luminosity function of galaxies into agreement with the predicted dark halo mass function \citep{Conroy2009,Moster2013}.

In the mass regime of dwarf galaxies ($M_* < 10^{10} M_\odot$), a large body of theoretical work attributes the regulation of star formation solely to reionization and stellar feedback, by means of radiation from young stars and supernova explosions \citep{Benson2002a,Bower2006}. In fact, powerful outflows driven by star formation have been observed in starbursting dwarfs such as M82 \citep{Martin_1998,Strickland_Stevens_2000}. Further observational evidence suggests that supernova feedback is likely to dominate in dwarfs \citep{Navarro2018}.  Quenched dwarf galaxies are rare, and the vast majority of them are found within 4 virial radii of a larger galaxy \citep[$M_K < -23$;][]{geha2012}, so environmental effects are likely largely responsible for gas removal from low mass galaxies.  However, a small number of isolated dwarf galaxies are observed with no signs of recent star formation \citep{Janz2017}, showing that processes internal to dwarf galaxies cannot be ruled out as quenching mechanisms.  Interestingly, evidence of AGN via optical and infrared (IR) indicators has been detected in hundreds of nearby dwarf galaxies \citep{Reines2013, Moran2014, sartori2015}.  \cite{Kaviraj2019} report the IR-selected AGN occupation fraction in high mass galaxies to be $1-3\%$, while the same criteria yield a $10-30\%$ fraction in dwarf galaxies ($M_* \sim 10^{8-10} M_\odot$).  Given that there are several factors that hinder the detection of AGN in dwarfs \citep{Satyapal2018, Cann2019}, this large AGN fraction can be regarded as a lower limit.  These studies suggest that AGN are common and potentially important phenomena in the low mass regime.  

The role of AGN feedback in dwarf galaxy evolution is being explored theoretically, with diverse results.  In idealized environments and using simplifying assumptions, such as spherical symmetry and single-phase ISM, \citet{Dashyan2017} find that AGN can be more efficient than star formation at clearing dwarf galaxies of their gas.  \cite{Koudmani2019} explore a range of photoionization models and outflow geometries in isolated dwarfs, concluding that AGN are unlikely to regulate star formation but can boost the energetics of existing stellar-powered outflows.  Cosmological simulations considering AGN feedback in high-$z$ dwarf galaxies indicate that BHs can provide a significant amount of feedback, able to effectively quench star formation \citep{Barai2018} and potentially reconcile observed dwarf galaxy anomalies with $\Lambda$CDM predictions \citep{Silk2017}.  Conversely, \cite{Trebitsch2018} find that even in the most extreme BH growth scenario, AGN feedback is negligible in comparison to ionizing starlight. In fact, they find that supernovae feedback stunts BH growth and quenches AGN feedback.  This wide variation in theoretical results demonstrates the need for observational constraints on the coupling of AGN energy to the interstellar medium (ISM) of dwarf galaxies.  

Encouragingly, observational evidence of AGN feedback in dwarf galaxies is beginning to surface.  In a study of 69 quenched dwarf galaxies from the MANGA survey \citep{Bundy2015}, \cite{Penny2018} find hints of a correlation between low star formation activity and the presence of AGN.  Six quenched galaxies in their sample show signs of AGN, and five of their six AGN candidates show ionized gas kinematically decoupled from their stellar velocity fields, suggestive of either outflows or recently accreted gas.  Additionally, \cite{Bradford2018} have found gas-depleted isolated dwarf galaxies with optical line ratios consistent with AGN, and \cite{Dickey2019} report AGN-like line ratios in a majority of the quiescent galaxies in their sample.

These promising results challenge current conceptions of feedback in dwarf galaxies and raise the question of whether the gas escapes the dark matter halo, permanently exhausting star formation in the galaxy, or stalls in the halo to cool and fall back later as part of a cycle of active and quiescent phases in a dwarf galaxy's life.  This will ultimately depend on the mass removed by the outflow and its relative velocity compared to the velocity needed to escape the combined gravitational pull of the dwarf galaxy and its dark matter halo.  

Additionally, dwarfs are expected to have a lower merger rate and a more quiet merger history than more massive galaxies \citep{Rodriguez-Gomez2015}, making their BHs closer relics of the initial BH seeds from which they originated (e.g., \cite{vanWassenhove2010}).  If AGN feedback influences the growth of BHs in dwarf galaxies, these objects will not be useful tracers of the seed BH population, and will have significant implications for BH seed formation models \citep{Mezcua2019}. 

A complete understanding of AGN feedback will rely on thorough investigations of outflows affecting all phases of the interstellar medium (e.g. \citep{Cicone2018}).  In the meantime, direct detection and measurement of any AGN-driven outflows in dwarf galaxies will be a crucial first step.  In this work, we examine a sample of mostly isolated dwarf galaxies, both with and without optical and IR signs of AGN activity.  We directly detect, measure and characterize ionized gas outflows in 13 dwarf galaxies and report the first velocity measurements of AGN-driven outflows in this mass regime.

We present our sample selection and observations in Section~\ref{sec:data}.  We describe our fitting methods and kinematic analysis in Section~\ref{sec:analysis}, with the criteria for outflow detection presented in Section~\ref{sec:outflow_detection}.  We present the results of our fits in Section~\ref{sec:results}, provide our physical interpretation of our results in Section~\ref{sec:discussion}, and summarize in Section~\ref{sec:summary}.  Throughout the paper we assume the cosmological model $H_0 = 71\,\rm{km}\,\rm{s}^{-1}\,\rm{Mpc}^{-1}$, $\Omega_m = 0.27$, and $\Omega_\Lambda = 0.73$.

\section{Data} \label{sec:data}

\subsection{Sample Selection} \label{sec:sample}
Drawing from the Sloan Digital Sky Survey Data Releases 7 and 8 (SDSS DR7,8), \cite{Reines2013}, \cite{Moran2014}, and \cite{sartori2015} (hereafter RGG13, M14, and S15, respectively) have identified hundreds of nearby ($z < 0.1$) dwarf ($M_* < 10^{10} M_\odot$) galaxies that exhibit optical and IR signatures of AGN. RGG13, M14, and S15 selected their samples using standard optical emission line diagnostics which probe the hardness of the dominant ionizing source. RGG13 identified 151 AGN candidates using the Baldwin, Phillips \& Terlevich (BPT) diagnostic \citep{baldwin1981} and the presence of broad H$\alpha$ emission lines.  M14 also drew from SDSS and used the BPT, \ion{He}{2}, [\ion{S}{2}], and [\ion{O}{1}] AGN indicators to build their sample.  They imposed further distance and photometry cuts in order to probe the low luminosity regime, resulting in a sample of 28 AGN. S15 extended the number of AGN candidates significantly (336) by including sources which have \ion{He}{2} emission and redder mid-IR colors consistent with AGN \citep{Jarrett2011, Stern2012}.  Due to their similar selection criteria, there is significant overlap between these three parent samples.

\begin{figure}
\includegraphics[width=\linewidth,clip]{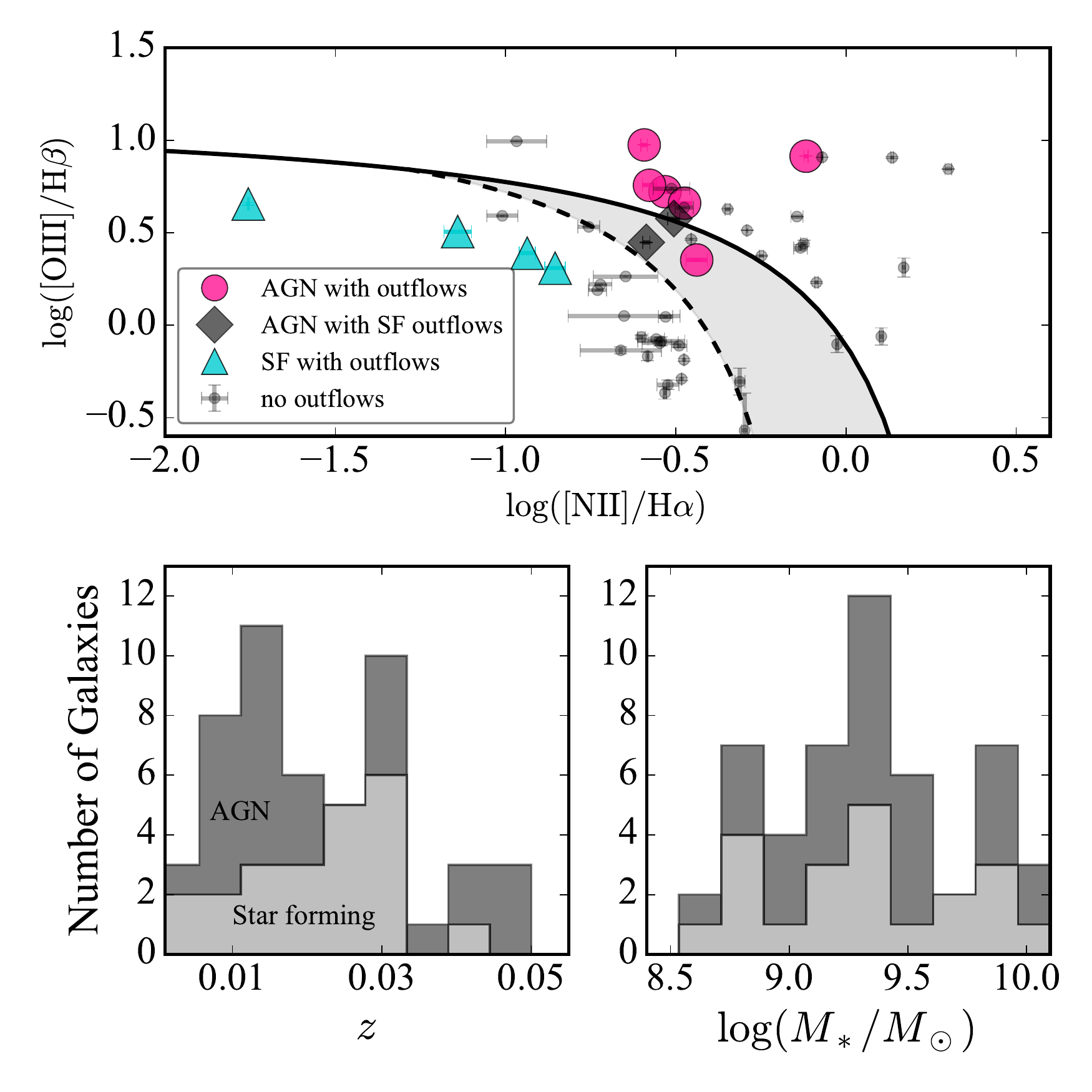}
\caption{Top: BPT line ratios from the central 0.2 kpc of each galaxy for dwarf galaxies in our sample observed with Keck/LRIS. 29 galaxies with optical signatures of AGN were selected from the samples of RGG13, M14, and S15.
The remaining 21 are a control sample composed of star-forming dwarf galaxies.  Dwarfs that present signatures of gas outflows are plotted with larger symbols. Bottom: Redshift and stellar mass distribution of our sample of dwarf galaxies.}
\label{fig:bpt_hist}
\end{figure}

From the parent samples of RGG13, M14, and S15, we selected candidate AGN host galaxies for follow-up study based on (1) emission line ratios falling above the star forming sequence \citep{kauffmann2003} on the BPT diagram and (2) presence of the high ionization He\,II emission line \citep{shirazi2012}.   Priority was given to spatially extended galaxies, and face-on galaxies were excluded whenever possible, to facilitate spatially resolved kinematic and emission line diagnostic measurements.  Other than this, galaxies were selected from the subsample based on the right-ascensions that could be observed during our observing time.   The resulting observed sample contained 29 galaxies with optical signatures of AGN activity (Fig.~\ref{fig:bpt_hist}). Fifteen of the AGN in our sample were selected from RGG13, an additional nine only from M14, and five more only from S15.

A control sample of galaxies with no optical or IR signatures of AGN was selected from the OSSY catalog \citep{OSSY2011}, which provides spectral line analysis for the entire SDSS DR7 atlas. We matched the sample with the MPA-JHU catalog \citep{Brinchmann2004, kauffmann2003, tremonti2004}, in order to obtain stellar masses. We applied the same redshift and stellar mass cuts as the AGN sample and excluded all galaxies falling above the star forming sequence on the BPT diagram, as well as all galaxies with detectable He\,II \citep{shirazi2012}. We also excluded potential AGN based on the WISE mid-IR color criteria \citep{Jarrett2011, Stern2012}. From this parent sample, we selected spatially extended galaxies to match the characteristics of the AGN sample.  NGC\,1569 was included specifically to facilitate comparison with previous kinematic 
studies \citep{Martin1998}.
The resulting sample of 21 star-forming dwarf galaxies were those with coordinates that could be observed during the Keck observing runs.  BPT line ratios from the central 0.2 kpc and corresponding redshift and mass distributions of our targets are shown in Fig.~\ref{fig:bpt_hist}. The full sample is described in detail in Manzano-King, et al. (in prep). 

Note that the control sample could potentially contain AGN that elude the selection criteria we used. Several factors can hinder the detection of AGN in dwarf galaxies, including dilution by the host galaxy (e.g. \citet{Moran2002}), low metallicity \citep{Kewley2008}, and AGN variability \citep{Lintott2009}.  In each case, the qualitative effect is to obscure existing AGN signifiers.

\begin{figure}
\includegraphics[width=\linewidth,clip]{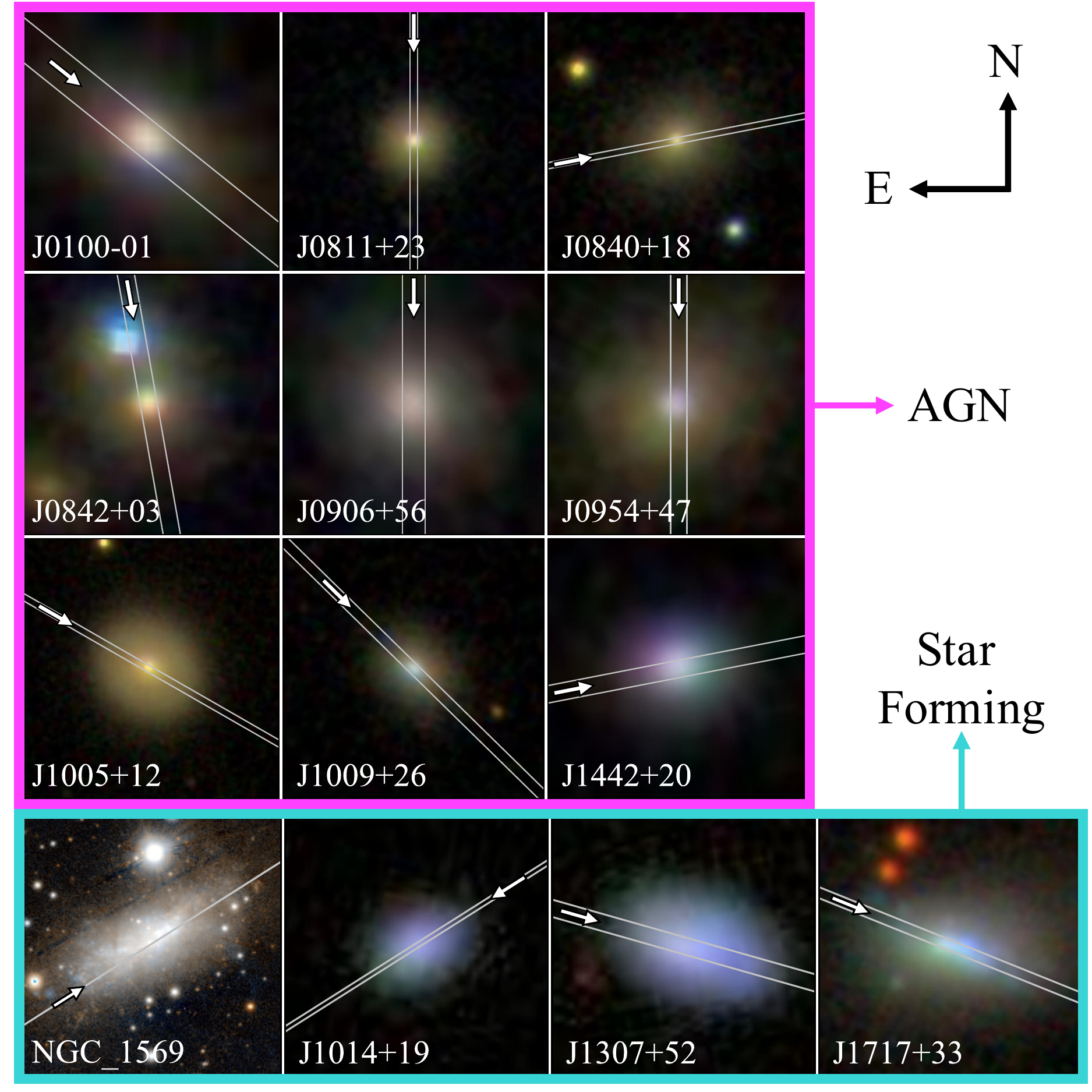}
\caption{SDSS color images of the 13 dwarf galaxies with spatially extended outflows.  All images were generated using the SDSS DR12 finding chart tool, with the exception of NGC 1569, which is outside of the SDSS footprint.  The NGC 1569 thumbnail is a PanSTARS z,g band color image rendered in the Aladin Lite Viewer with a 3' field of view ($\sim 1\,$kpc on a side).  Each SDSS image is scaled to 10 kpc on a side and the placement of the 1 arcsecond-wide slit is shown in light gray.  
}
\label{fig:thumbs}
\end{figure}

\subsection{Observations and Data Reduction}
\label{sec:obs}
We obtained longslit spectroscopy of the 50 targets using the Low Resolution Imaging Spectrometer \citep[LRIS;][]{Oke_1995, Rockosi2010} on the Keck I telescope on the nights of UT 2015 March 24--25, December 4--5, and 2017 June 24--25.  We placed a 1 arcsecond-wide slit along the semimajor axis (see Fig.~\ref{fig:thumbs}) of each galaxy, projecting to $\sim$ 7 pixels on both the blue and red CCDs (Marconi on LRIS-B and LBNL on LRIS-R).   The objects in our sample are located at distances between 3 and 209 Mpc  ($z < 0.05$), which yield spatial scales between 14 and 965 pc arcsec$^{-1}$ \citep{Wright2006}.    For the blue side (LRIS-B), we used the 600 groove mm$^{-1}$ grism blazed at 4000\AA, yielding a dispersion of 0.63\AA\,pixel$^{-1}$.  For the red side we used either the 600 groove mm$^{-1}$ grating blazed at 5000\AA, the 900 groove mm$^{-1}$ grating blazed at 5500\AA, with a 5600\AA\ dichroic, or the 1200 groove mm$^{-1}$ grating blazed at 7500\AA, with a 5000\AA\ dichroic, yielding dispersions of 0.80, 0.53, and 0.40\AA\,pixel$^{-1}$, respectively.  We obtained one or two 1200 second exposures of each galaxy.  To achieve the signal to noise ratio necessary for proper fitting of the spectra, a second exposure was necessary for galaxies with surface brightness $\gtrapprox21.5\,\text{mag\,arcsec}^{-2}$.
 The weather was clear on all six nights, and the seeing was typically under 0$\farcs$6, except on UT 2015 March 24, when it was 0\farcs7.  For details of observations of individual objects see Manzano-King et al.\ (in prep).
The data were reduced using a Python pipeline to automate the standard IRAF reduction tasks.  
Flexure on the red camera was corrected using the average shift in sky lines.  There are not enough sky lines to correct for flexure on the blue CCD, so each galaxy spectrum on the blue side was redshift-corrected using the redshift measured from the flexure-corrected red spectrum.  Flexure on the blue CCD was then calculated by comparing galaxy emission lines with their expected rest frame values. 
The longslit spectra were rectified along both the wavelength and spatial axes, creating 2-dimensional spectra where each pixel row is a fully reduced 1-dimensional spectrum.  Customized extractions along the slit are obtained by summing pixel rows, allowing us to preserve spatial information by extracting small sections of the spectra or to sum over large apertures when higher signal to noise is required.
%
\section{Analysis}
\label{sec:analysis}
Visual inspection of the emission line profiles of the galaxies in our sample revealed clear broadened, often blueshifted wings in some of the galaxies.  Such line profiles are often indicative of gas outflows.   Thus, in order to systematically search for outflows in our sample, we conducted a detailed and consistent fitting process as described below.

\subsection{Fitting the Stellar Continuum}
\label{sec:ppxf}

For each of the 50 galaxies in our extended sample, we used the same fitting algorithm to systematically identify and measure outflows.  For each source, we began by extracting a one-dimensional spectrum of the entire galaxy, when possible.  Some galaxies in our sample have contaminating foreground stars, in which case we chose custom apertures to maximize the extraction region while excluding the contaminating sources.  

In order to measure accurate emission line fluxes, it is necessary to account for stellar absorption, which primarily affects the Balmer emission lines. Penalized Pixel-Fitting method (pPXF) \citep{Cappellari2017} is an algorithm designed to extract stellar kinematics using a maximum penalized likelihood approach to match stellar spectral templates to absorption features in galaxy spectra.  In order to minimize template mismatch, stellar templates with relevant stellar types, negligible emission line contamination, and full wavelength coverage for the wavelength range in our galaxy spectra were chosen from the Indo-US Library of Coud{\'e} Feed Stellar Spectra \citep{Valdes2004}.  pPXF is designed to choose the appropriate stellar templates automatically, but it is necessary to check that emission line contamination from the ISM in the templates does not skew the fit, ultimately resulting in underestimation of Balmer emission line fluxes.  Visual inspection of the fit to features such as the Ca\,II $\lambda$3933 line, which is free from contaminating emission lines, allows us to assess whether the stellar continuum and absorption features are being properly estimated.  An example of a fit to the stellar spectrum of a galaxy is shown on the top panel of Fig.~\ref{fig:OIII_fit}.  Careful selection of stellar templates is especially important for young stellar populations, whose fits are often much improved by considering only B and A-type stellar templates in order to carefully fit their prominent Balmer absorption features.

\subsection{Fitting Emission Lines}
\label{sec:fitting}
%

\begin{figure}
\includegraphics[width=\linewidth,clip]{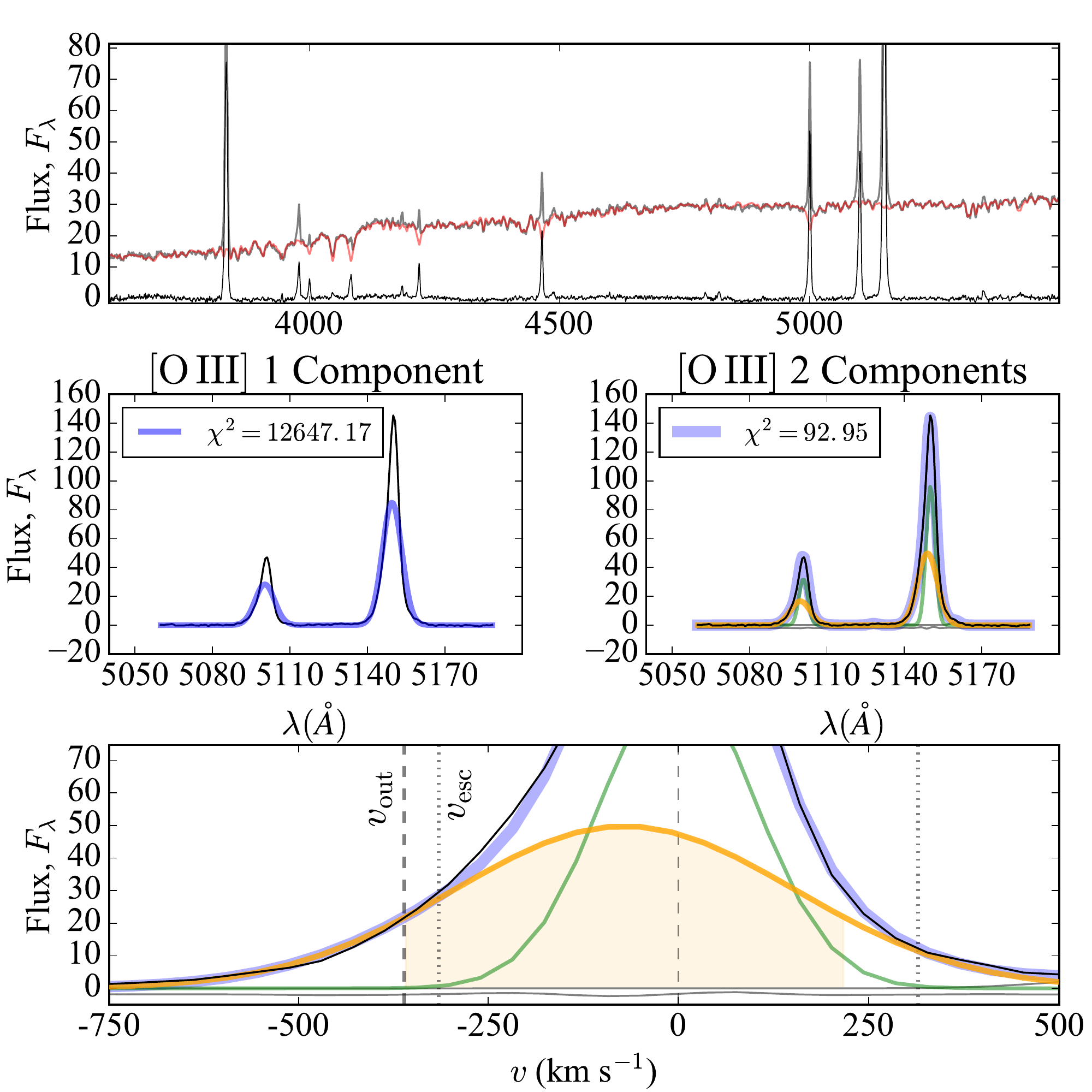}
\caption{Top: The spectrum of J084234.51+031930.7 is extracted from within $R_{50}$ (gray).  The pPXF best fit stellar continuum (red) is subtracted, leaving the residual emission spectrum (black).  The flux units in this figure are $(\text{erg\,cm}^{-2}\,\text{s}^{-1}\,$\mbox{\AA}$^{-1})$.
Middle: One- and two-component Gaussian fits to the [\ion{O}{3}] doublet are shown.  The one-component model on the left is clearly a poor fit compared to the multicomponent fit on the right.  The narrow (green) and broad (orange) component color scheme is used throughout this paper. 
Bottom: A close up of the outflow component of [\ion{O}{3}]$\lambda$5007 and its $W_{80}\sim\text{FWHM}$ shaded in orange, indicating the outflow velocity (dashed line) and escape velocity of this galaxy (dotted line).}
\label{fig:OIII_fit}
\end{figure}

After subtracting the best-fitting stellar population model of the galaxy, the residual emission lines were fit using a custom Bayesian MCMC maximum likelihood sampling algorithm, implemented using the Python package {\it emcee} \citep{emcee}.  Due to their strength and general isolation from other strong spectral features, we used the [\ion{O}{3}] $\lambda\lambda$4959,5007 doublet to constrain line profiles for the rest of the emission lines in the spectra.  The middle panels of Fig.~\ref{fig:OIII_fit} show that a single Gaussian produces a poor fit when broadened wings are visible in the [\ion{O}{3}] doublet.  To address this, we constructed a double-Gaussian model consisting of a narrow and a broad component.  In order to reduce the number of free parameters in the [\ion{O}{3}] line profile fit, we placed three constraints on each Gaussian component: (1) Based on transition probabilities, the flux ratio between [\ion{O}{3}]$\lambda$4959 and [\ion{O}{3}]$\lambda$5007 was fixed to 1:3; (2) the velocity widths derived from each Gaussian component were forced to be equal in both emission lines in the doublet; (3) the spacing between the emission lines was held fixed at 2869.35 km\,s$^{-1}$ or 47.92\AA.  \ion{Fe}{1}\,$\lambda$4985,\,4986,\,5016 lines contribute a small amount of additional flux which causes the fit to overestimate the width of the broad component.  To address this contamination, we model them with independent amplitudes and widths equal to that of the narrow component of [\ion{O}{3}].  The following were left as free parameters: (1,2) Amplitudes of narrow and broad components; (3,4) velocity widths of narrow and broad components; (5) recession velocity of the doublet, set by the mean velocity of the narrow component of [\ion{O}{3}]$\lambda 5007$; (6) velocity offset between narrow and broad components; and (7-9) amplitudes of contaminating \ion{Fe}{1} lines.

\subsection{Detecting Outflows}
\label{sec:outflow_detection}
Broadened or shifted components in emission lines trace gas with different kinematics than the rest of the ionized gas in the galaxy.  Such components potentially trace outflowing gas.  In order to systematically detect outflows, we fit both single and double Gaussian models to the [\ion{O}{3}] doublet in each of the 50 galaxies in our sample.  Fits were performed on one-dimensional spectral extractions covering as much of each galaxy as possible, as described in Section~\ref{sec:ppxf}.  Details of the double Gaussian models are described in Section~\ref{sec:fitting} and examples of the single- and double-component fits are shown in the two middle panels of Fig.~\ref{fig:OIII_fit}.

An additional Gaussian component always reduced the $\chi^2$ of the fits, so a rigid goodness-of-fit threshold to determine whether a line requires a second Gaussian would not yield robust results.  Instead, we flagged each fit that produced a second kinematic component wider than the instrumental resolution ($\sigma > 80\, \rm{km\,s}^{-1}$) and amplitude higher than the noise level of the spectrum.  This approach yielded a subsample of 15 candidate outflow galaxies.  

Next, we examined the spatial extent of these broad components by repeating the double Gaussian fits on smaller extraction regions along the slit in each of the 15 galaxies.  We found that the broad [\ion{O}{3}] components in two AGN-hosting galaxies, J094941.20+321315.9 and J141208.47+102953.8 are spatially unresolved.  The remaining 13 galaxies, however, show clear broadened components in their [\ion{O}{3}] at all radii, as long as the emission line was detectable (typically out to one half light radius, or $\sim 1-3$ kpc).  Based on their narrow component emission line ratios, nine of these 13 galaxies with spatially resolved broad components are classified as AGN and four are star forming (SF).  Full names, redshifts, stellar masses, and narrow line BPT classifications for these 13 outflow galaxies are listed in Table~\ref{table:outflow}.

Of our full sample of 50 galaxies, 31\% (9/29) of the AGN and 19\% (4/21) of the SF galaxies show spatially extended outflows.  If we include the two AGN with spatially unresolved broad components, we observe outflows in 38\% of the AGN in our sample.

\subsection{Outflow Velocity}
\label{sec:kinematics}
Table~\ref{table:outflow} lists fit outputs to one-dimensional spectrum extractions within the r-band Petrosian half light radius (hereafter $R_{50}$) of each of the 13 galaxies discussed in this work. $W_{80}$, the velocity width containing 80\% of the flux of an emission line, is a widely used metric to describe outflow velocities \citep{Harrison2014}.  For a single Gaussian profile, $W_{80} = 1.09\, $FWHM.  We define the outflow velocity 

\begin{equation}
v_{\rm out} = -v_0 + \frac{W_{80}}{2}
\end{equation}

where $v_0$ is the velocity offset between the center of the narrow and broad components (see Fig.~\ref{fig:OIII_fit}) and is negative when blueshifted.  


\subsection{Decomposed Flux Ratios}
\label{sec:ratios}

\begin{figure*}
\includegraphics[width=\linewidth,clip]{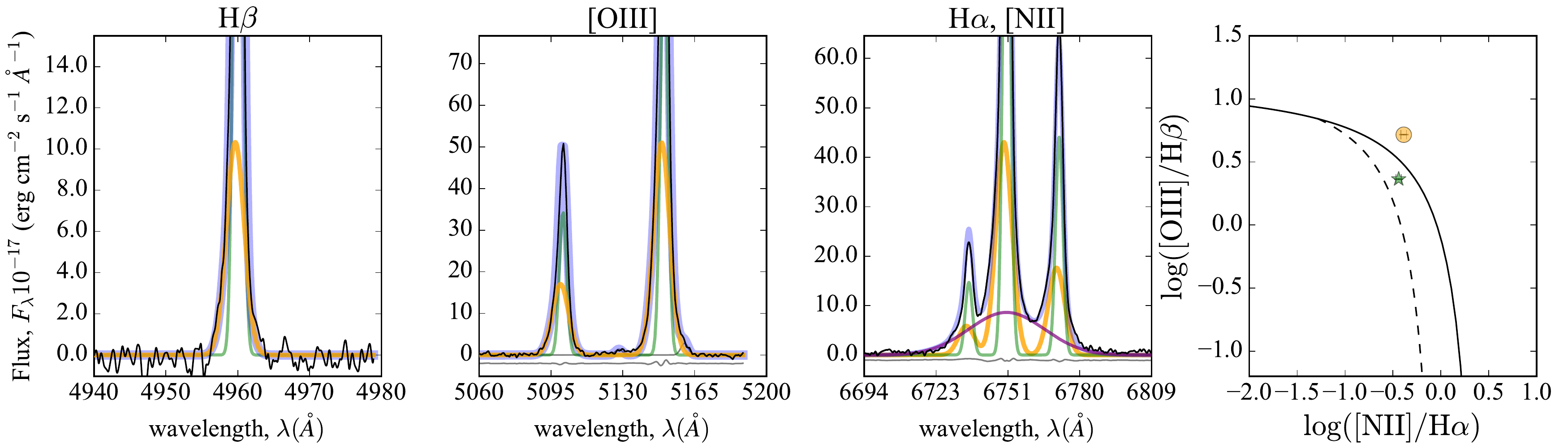}
\caption{Emission lines for J084234.51+031930.7 are shown as an example of how the kinematic components can be decomposed and placed on the BPT diagram.  As described in Section~\ref{sec:fitting} and illustrated in Fig.~\ref{fig:OIII_fit}, the stellar continuum is subtracted and the residual [\ion{O}{3}] doublet is fit using a double Gaussian model which includes contaminating \ion{Fe}{1} lines.  The widths of H$\alpha$, H$\beta$, and [\ion{N}{2}] are fixed based on the [\ion{O}{3}] model, but the fluxes for each kinematic component are left as free parameters.  The resulting line flux ratios for the narrow (green star) and the broad component (orange circle) are plotted on the BPT diagram (right). In this case, the model required an additional broad H$\alpha$ component, which was used to obtain a black hole mass (See Appendix~\ref{sec:M_bh}).}
\label{fig:fits}
\end{figure*}

In most of the galaxies discussed here, the broad emission line component can be detected in several emission lines.  When possible, we applied the two-component  kinematic [\ion{O}{3}] model to H$\alpha$, H$\beta$, and[\ion{N}{2}] in order to place the outflowing gas on the BPT diagram.  The kinematic model  (velocity offsets and widths) measured in [\ion{O}{3}] are held fixed while the amplitudes of each kinematic component are allowed to vary independently.  Due to their small separations, H$\alpha$ and the [\ion{N}{2}] doublet were fit simultaneously, with the flux ratio of [\ion{N}{2}]$\lambda$6548 and [\ion{N}{2}]$\lambda$6563 fixed at 1:3 in accordance with their transition probabilities.  Two galaxies (J084234.51+031930.7 and J100935.66+265648.9) required an additional broad H$\alpha$ component in order to constrain the outflow  H$\alpha\,/$[\ion{N}{2}] flux ratio.  See Appendix \ref{sec:M_bh} for further discussion on additional broad H$\alpha$ components associated with the broad line region (BLR).  The full multicomponent fits to H$\beta$, [\ion{O}{3}], H$\alpha$, and [\ion{N}{2}] as well as the resulting kinematically decomposed BPT ratios for J084234.51+031930.7 are shown in Fig.~\ref{fig:fits}.  


\begin{figure}
\includegraphics[width=\linewidth,clip]{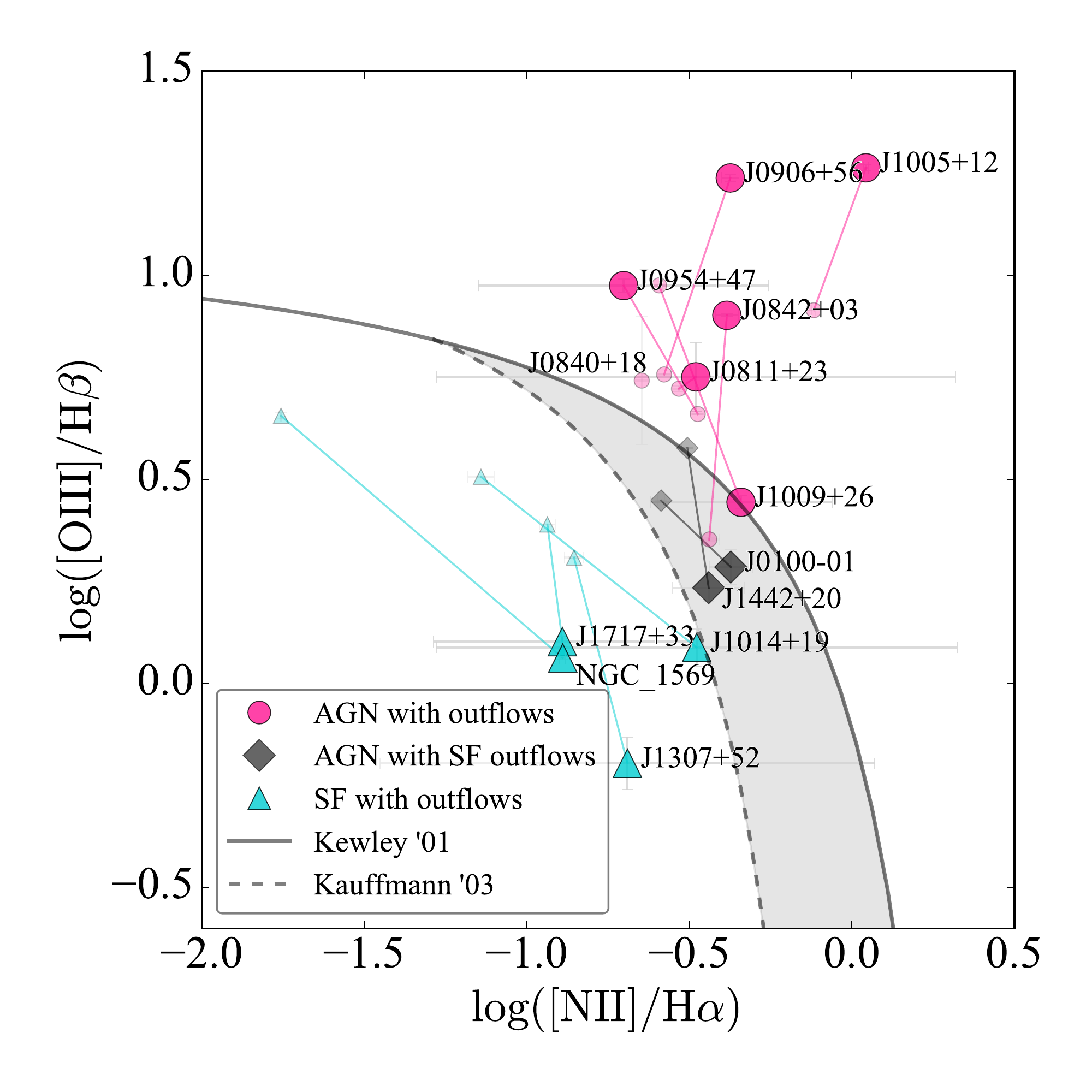}
\caption{Narrow and broad emission line ratios are shown for each galaxy with outflows.  The smaller symbols show the position of the bound gas (narrow component in spectrum) for each galaxy, and the larger symbols those of the outflow (broad component).  In this paper, each galaxy is classified based on its position of the outflow component in the AGN (pink circles), composite (gray diamonds), or star forming regions (cyan triangles).}
\label{fig:bpt_broad}
\end{figure}

\section{Results}
\label{sec:results}


\subsection{Classifying Outflows}
\label{sec:class}
%

In Fig.~\ref{fig:bpt_broad}, we show the position of the narrow and broad components (small and large symbols, respectively) on the BPT diagram, obtained following the procedure described in Section~\ref{sec:kinematics}.  The two components for each galaxy are connected by a thin line of the same color.  

The star forming galaxies (cyan triangles) were included in our control sample due to their narrow components' positions on the BPT star forming sequence.  Two of the SF objects (J101440.21+192448.9 and J130724.63+523715.2) are from the RGG13 
sample of BPT star forming galaxies with broad H$\alpha$ lines.  Follow-up observations of these two objects found that these broad H$\alpha$ features were transient and likely due to Type II supernovae \citep[][also confirmed by our observations]{baldassare2016}. The outflows in all four galaxies fall within the star forming region of the BPT diagram.

Emission line ratios falling above the \cite{kauffmann2003} (dotted) line are widely regarded as evidence of AGN activity.  However, a well known limitation of the BPT diagram in definitively identifying AGN is that shock ionization with significant contribution from precursor gas can produce line ratios in star forming galaxies that mimic those of AGN \citep{Allen2008}.  Strong near IR detections of the [\ion{Si}{6}] coronal line provide independent confirmation of the presence of AGN in four of the dwarfs with outflows (J090613.75+561015.5, J095418.16+471725.1, J100551.19+125740.6, and J100935.66+265648.9; Bohn et al., in prep).  Near IR line ratio diagnostics confidently exclude shocks as the originating ionizing mechanism in these objects.  The [\ion{Ne}{5}]$\lambda$3426 coronal line is also detected in these four AGN as well as J084025.54+181858.9.  These measurements provide additional evidence for the presence of AGN at least in these five objects.  In this work, we classify the nine objects with narrow emission lines above the dotted \cite{kauffmann2003} line as AGN.

Six dwarfs with AGN have broad components that occupy the region above the \cite{kewley2001} maximum starburst line (solid) on the BPT diagram, so we plot them as pink circles throughout this paper.  J084025.54+181858.9 did not have a sufficiently high signal-to-noise ratio to fit the outflow in lines other than [\ion{O}{3}], so we count this object among the AGN-driven outflow sample based on its AGN-consistent narrow emission line ratios.  Two of the AGN (gray diamonds) have outflow components that fall in the composite region, indicating a significant contribution from star formation.  

Throughout this paper, we classify each outflow based on position of the broad component on the BPT diagram.  Table~\ref{table:outflow} is sectioned according to the broad component BPT classification and lists the BPT classification for both narrow and broad components in the last two columns.


\subsection{Integrated Properties of Outflows}
\label{sec:integrated_outflows}

\begin{figure*}
\includegraphics[width=\linewidth,clip]{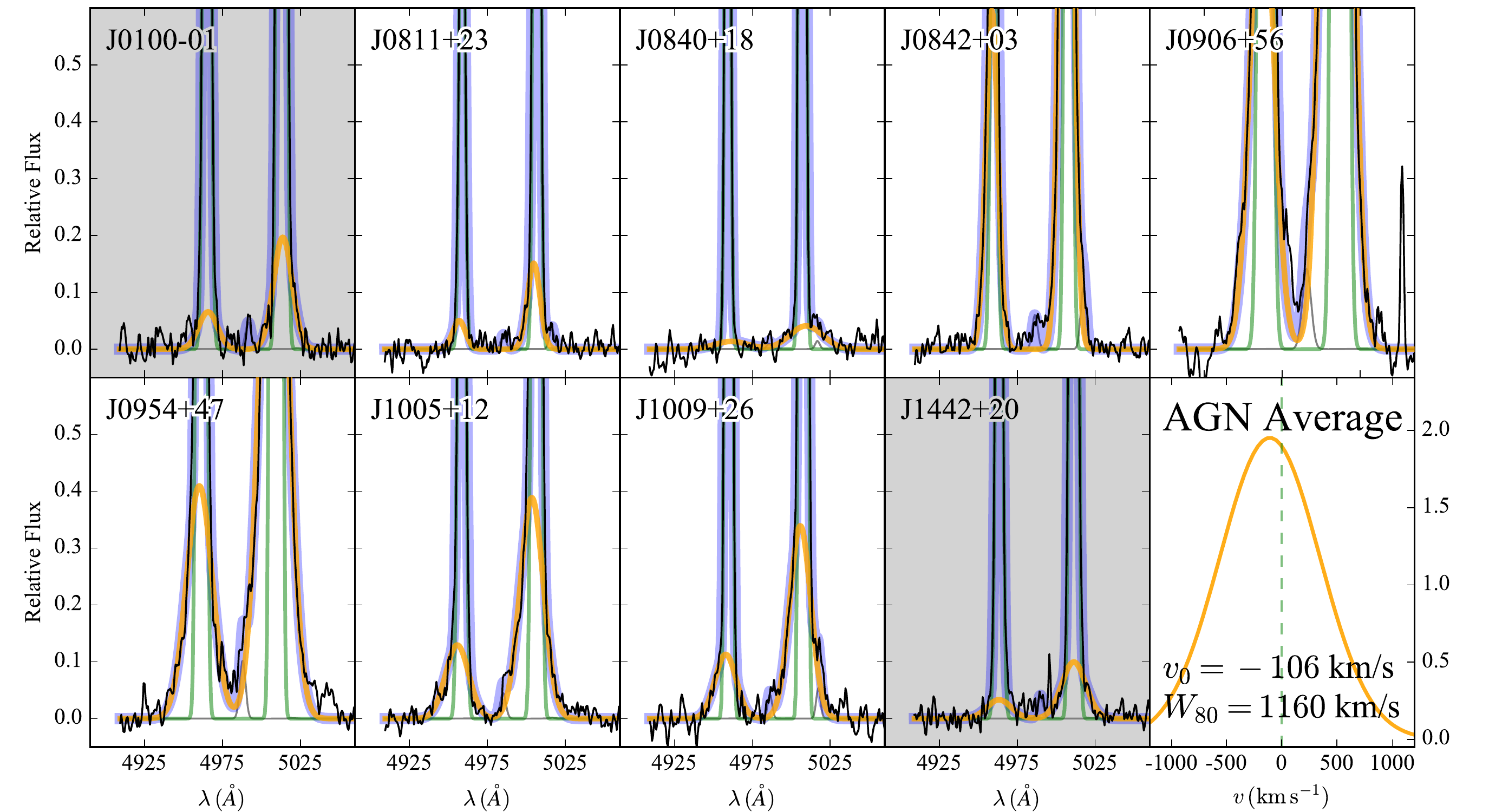}\\
\includegraphics[width=\linewidth,clip]{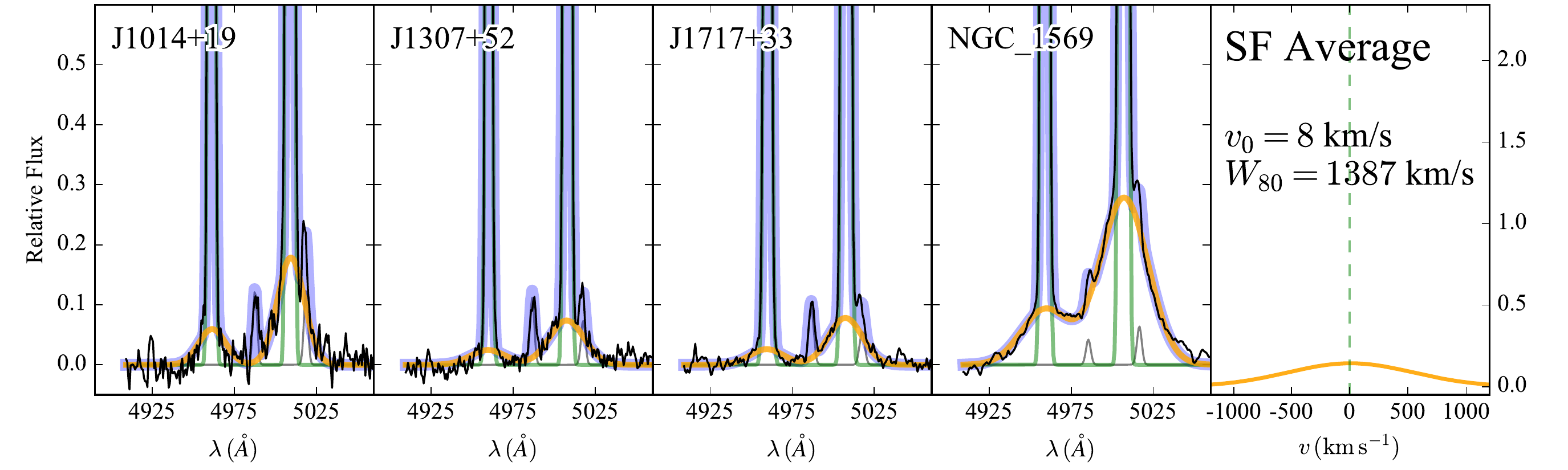}
\caption{Details of the broad component fits to the [\ion{O}{3}] doublets of each of the 13 galaxies with outflows are shown.  The spectra are extracted from the region within the $R_{50}$ of each galaxy and have all been normalized by the continuum flux just redward of [\ion{O}{3}]$\lambda$5007.  AGN are grouped in the top figure and star forming galaxies are on the bottom.  Average fit parameters, weighted by the luminosity of each narrow [\ion{O}{3}] line, are shown in the last panel of each section.  Panels shaded in gray are classified as composite and are excluded from the AGN outflow average.  Values of $v_0$, $W_{80}$, and $v_{\rm out}$ are listed for each of the 13 galaxies in Table~\ref{table:outflow}. 
}
\label{fig:stacked}
\end{figure*}

In Fig.~\ref{fig:stacked}, we illustrate the difference in line profiles between galaxies with and without AGN.  The galaxies are grouped according to the classification assigned in Section~\ref{sec:class}, with nine objects hosting AGN on top and four SF galaxies on the bottom.  Each panel shows multicomponent fits to the [\ion{O}{3}] doublet of each galaxy.  The spectra have been extracted from within $R_{50}$ and normalized by the continuum flux just redward of [\ion{O}{3}]$\lambda$5007.  As in Figs.~\ref{fig:OIII_fit} and \ref{fig:fits}, green curves trace the narrow component, orange curves indicate the outflow component, and contaminating Fe\,I lines are shown in gray.  The orange curves in the two bottom right panels represent averages of the fits to AGN and star forming outflows, weighted by the luminosity of each narrow [\ion{O}{3}] component.  The composite outflow objects are shaded gray and are not included in the weighted average for AGN.  

A comparison of the average AGN vs.\ star-forming emission line profiles reveals a fundamentally different line shape, with AGN outflows that are blueshifted with respect to the narrow component by  $\sim$100 km s$^{-1}$, on average. This is in contrast with the outflows in star forming galaxies that show virtually no velocity offset.  The broad component of the average AGN driven outflow line profile carries a larger percentage of the total [\ion{O}{3}] flux  than its star forming counterpart ($22.8\%$ and $4.9\%$, respectively).  

There is no discernible difference between the stellar masses of the galaxies with and without outflows.  However, the  six  galaxies  with  non-AGN  driven  outflows  show  signs  of very active star formation relative to the rest of the sample.  As we will further discuss in Section \ref{sec:feedback}, the galaxies with AGN-driven outflows have redder colors than the four star forming galaxies and two AGN with stellar driven winds.  These six  galaxies  with  non-AGN-driven outflows have the highest specific star formation rates ($\log(\rm{sSFR\,\rm{Gyr}^{-1}}) > \text{-}0.5 \,$) of the  entire  sample  of 50 dwarfs.  The galaxies with AGN-driven outflows have sSFRs similar to the galaxies with no outflows \citep[-0.5 $< \log(\rm{sSFR\,\rm{Gyr}^{-1}}) <$ -3; ][]{Conroy2009gran}.

\subsection{Spatially Resolved Properties of Outflows}
\label{sec:spatial_outflows}

\begin{figure}
\includegraphics[width=\linewidth,clip]{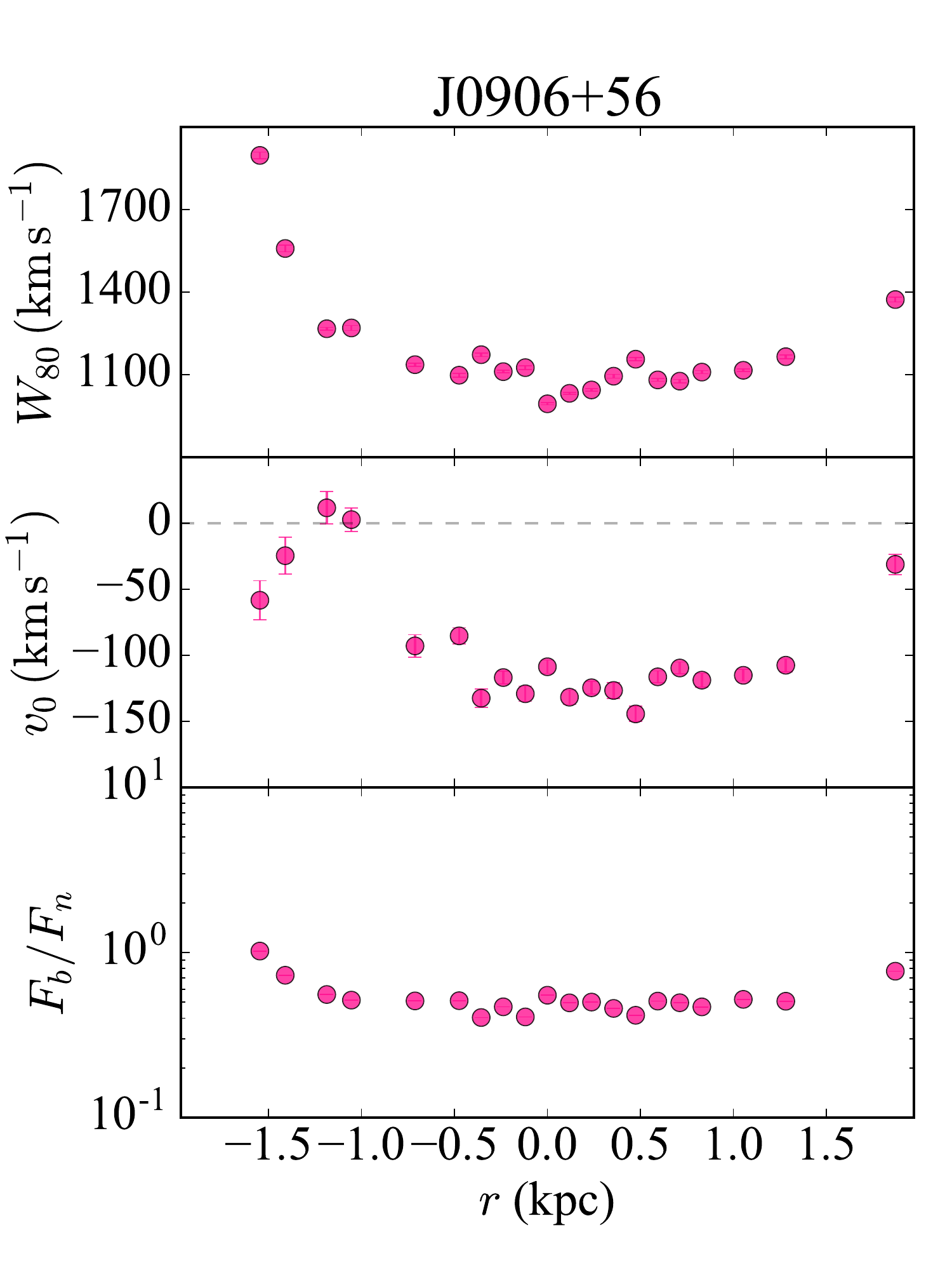}
\caption{Spatial properties of the AGN-driven outflow in J090613.75+561015.5.  The outflow width $W_{80}$ and offset $v_0$ as defined in Section~\ref{sec:kinematics} as a function of radius are shown in the top two panels, respectively.  The ratio of the broad to narrow [\ion{O}{3}] flux is shown in the bottom panel.
}
\label{fig:spatial}
\end{figure}

Common trends in the behavior of AGN-driven outflows as a function of radius are apparent.  In general, the width of the broad component ($W_{80}$) broadens, the mean velocity offset of the broad component relative to the narrow ($v_0$) approaches zero, and the ratio of the broad to narrow component fluxes $(F_b/F_n)$ increases with increasing radius.  These trends are typical for the AGN in our sample.  As an example, we show the spatial properties of the outflow in J090613.75+561015.5 in Fig.~\ref{fig:spatial}.  This object was chosen due to its intermediate mass ($2\times10^9M_\odot$) and color ($u-r_{model}\sim1.7$), with respect to the rest of the galaxies hosting AGN driven outflows discussed here. The detailed spatial properties of the remaining galaxies are shown in Appendix~\ref{apx:spatial}.

We confirm that the outflows extend all the way to the limit where our observations have a sufficiently high signal-to-noise ratio to fit [\ion{O}{3}] (between 1.5 and 3 kpc from the center).  In addition, their integrated line flux ratios are consistent with AGN out to these distances, indicating that the effect of the central AGN in these dwarfs is largely non-local.  

\citet{Greene2011} use spatially resolved longslit spectroscopy of luminous obscured quasars to extend the relation between NLR size and [\ion{O}{3}] luminosity and linewidth by an order of magnitude.  To test whether the AGN would be able to ionize gas out to the distances in their host galaxies where we detect signal, we estimate the expected sizes of the narrow line regions (NLR) using this relation.   We find that the NLR sizes range between 3.1 and 3.5 kpc, well beyond the distances where we are able to measure line ratios ($<$3 kpc in all cases).

\section{Discussion}
\label{sec:discussion}

\subsection{Ionization by Star Formation vs. AGN}
\label{sec:ionization}
%

For simplicity, the outflows in this paper are classified as SF, composite, and AGN-driven based on the position of their broad components on the BPT diagram in Fig.~\ref{fig:bpt_broad}.  

We see that the broad outflow components for all four SF galaxies (cyan triangles) fall below and to the right of their corresponding narrow components, likely due to the contribution from shock-induced photoionization \citep{Allen2008}.  Stellar-driven outflows are commonly observed in massive starburst galaxies \citep{Rupke2002} and, at least in those galaxies, shocks often propagate through the ISM \citep{Soto2012}.   Thus, it is likely that the fast outflows in these four star forming galaxies are driven by stellar processes and at least partially ionized by shock heating.  
Though we cannot positively rule out the presence of faint or extremely variable AGN, we find no optical or IR evidence of AGN that could be driving the outflows in these control sample galaxies.

Six objects in our sample have broad components falling on or above the \cite{kewley2001} line on the BPT diagram (solid black line in Fig.~\ref{fig:bpt_broad}), indicating that they exceed the theoretical maximum ionization possible with star formation alone.  We consider AGN to be the most likely mechanism for driving the outflows in these objects.

Two AGNs have outflow components that fall in the composite region, indicating a significant contribution from star formation to the ionization of these components.  This star formation contribution is more pronounced in the outflowing gas than the bulk of the bound gas (narrow component) in the galaxy.  Thus, we regard these two objects as dwarfs that likely contain AGN, but whose outflows might be at least partially powered by star formation.  

As mentioned above, we were not able to measure line flux ratios for the broad component of the remaining object, J084025.54+181858.9, due to their lower signal-to-noise ratio.

Note that J100935.66+265648.9 follows a similar trend to the two objects with composite broad components (gray diamonds) in that its broad component falls closer to the composite region of the BPT diagram relative to its narrow component. The inclusion of an additional broad H$\alpha$ component with FWHM $= 298\ \rm{km\,s}^{-1}$ 
to the H$\alpha$ $+$ [\ion{N}{2}] model described in Section~\ref{sec:fitting} was necessary to properly fit the outflow component fluxes for this object.  This additional component is too narrow to convincingly be associated with the broad line region, and is possibly tracing an additional kinematic component of the outflow or a velocity gradient along the slit introduced by rotation.  Though the broad emission line ratios appear to show significant dilution by star formation, the outflow remains on or above the \cite{kewley2001} maximum starburst line.  Thus, we count this object among those with AGN-driven outflows in our sample.


\subsection{Outflow Line Profiles}
\label{sec:geometry}
%

Additional evidence of outflow origin could be inferred from the different structure of the spectral lines for AGN vs.\ star formation powered outflows.  The average line profiles of AGN shown in Fig.~\ref{fig:stacked} are more blueshifted and slightly less broad than those of SF galaxies.  This could be interpreted by considering the physical position of the source of the winds.  In any given galaxy with radial outflows, the blueshifted gas will be easier to detect since the gas suffers from less galactic obscuration than the gas moving away from the observer \cite[e.g.,][]{Soto2012, Crenshaw2010}.   

We speculate that the blueshifted outflow component can be explained by considering obscuration from denser material in the central regions of the galaxy.  If there is only one source, and it is positioned in the center of the galaxy (as would be the case for AGN), most of the gas we observe will be blueshifted.  However, if there are multiple sources scattered around the galaxy, with some of them being on the near side of the galaxy, closer to the observer (as would be the case with supernovae and regions of star formation), then we will be more likely to observe both the blue- and redshifted emission.  

\begin{figure}
\includegraphics[width=\linewidth,clip]{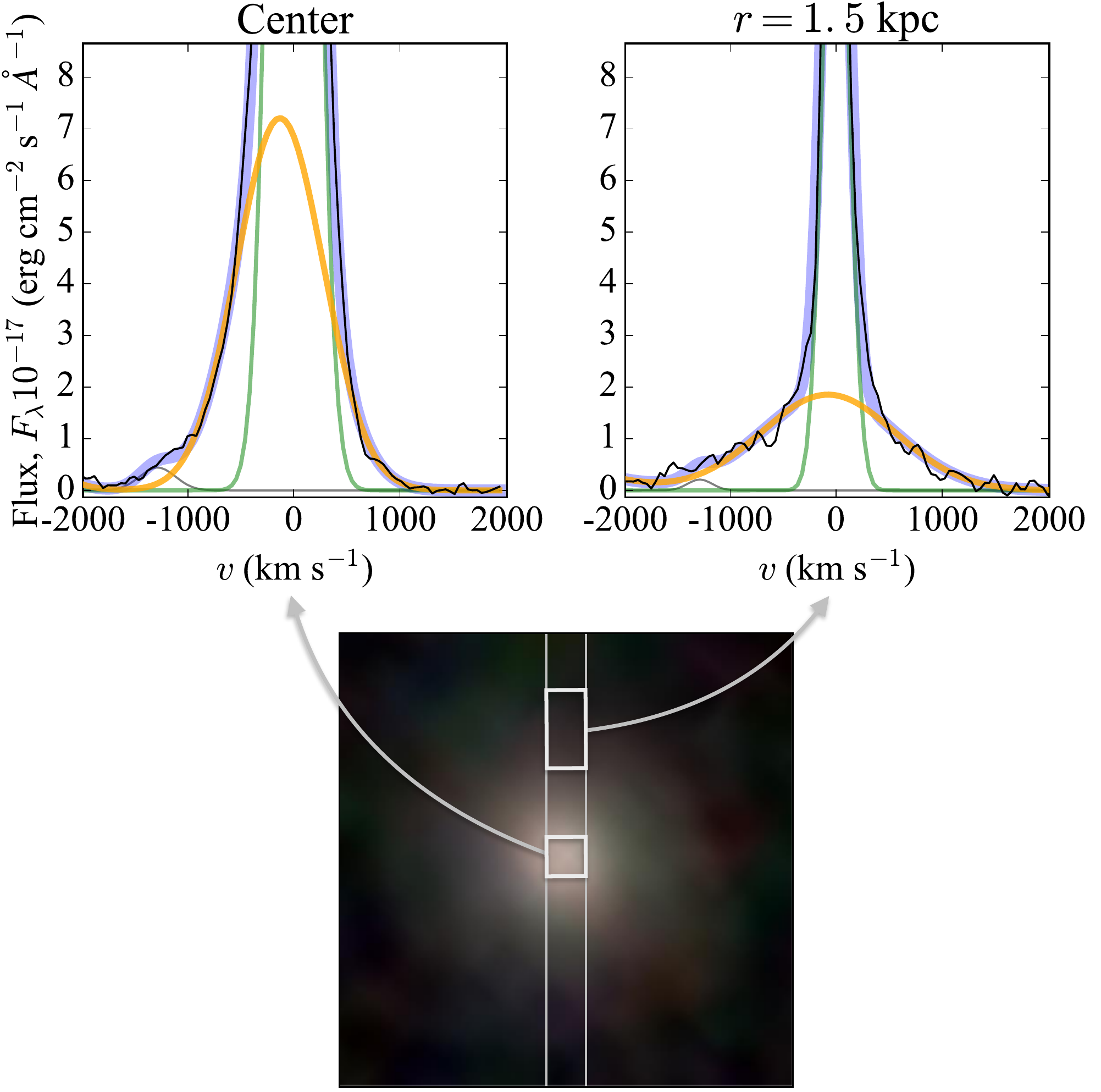}
\caption{A comparison between outflow component line profiles extracted from the central $\sim0.4$ kpc and $\sim1.5$ kpc from the center of J090613.75+561015.5 is shown.  We speculate that the central outflow appears blueshifted due to denser material in the center of the galaxy obscuring emission from the redshifted gas escaping the far side of the galaxy.  Farther from the center, the outflow profile widens and the velocity offset becomes less pronounced as galactic obscuration fades and both blue- and redshifted portions of the outflow is revealed.
}
\label{fig:spatial_thumb}
\end{figure}

In Fig.~\ref{fig:spatial_thumb}, we illustrate the difference in AGN-driven outflow line profiles extracted from the center and the outskirts of a galaxy, using J090613.75+561015.5 as an example.  The first panel shows the fit to [\ion{O}{3}]$\lambda$5007, extracted from the central $\sim0.4$ kpc-- note the blueshifted broad component.  An extraction $\sim1.5$ kpc from the center reveals a broad component with a smaller velocity offset and a slightly wider profile, consistent with the scenario where obscuration decreases with radius, revealing emission from both approaching and receding material. The typical radial trends of AGN-driven outflow properties introduced in Section \ref{sec:spatial_outflows} and Fig.~\ref{fig:spatial} would be a natural consequence of the proposed model.

\subsection{Gas Velocity}
\label{sec:velocity}
%
\begin{figure}
\includegraphics[width=\linewidth,clip]{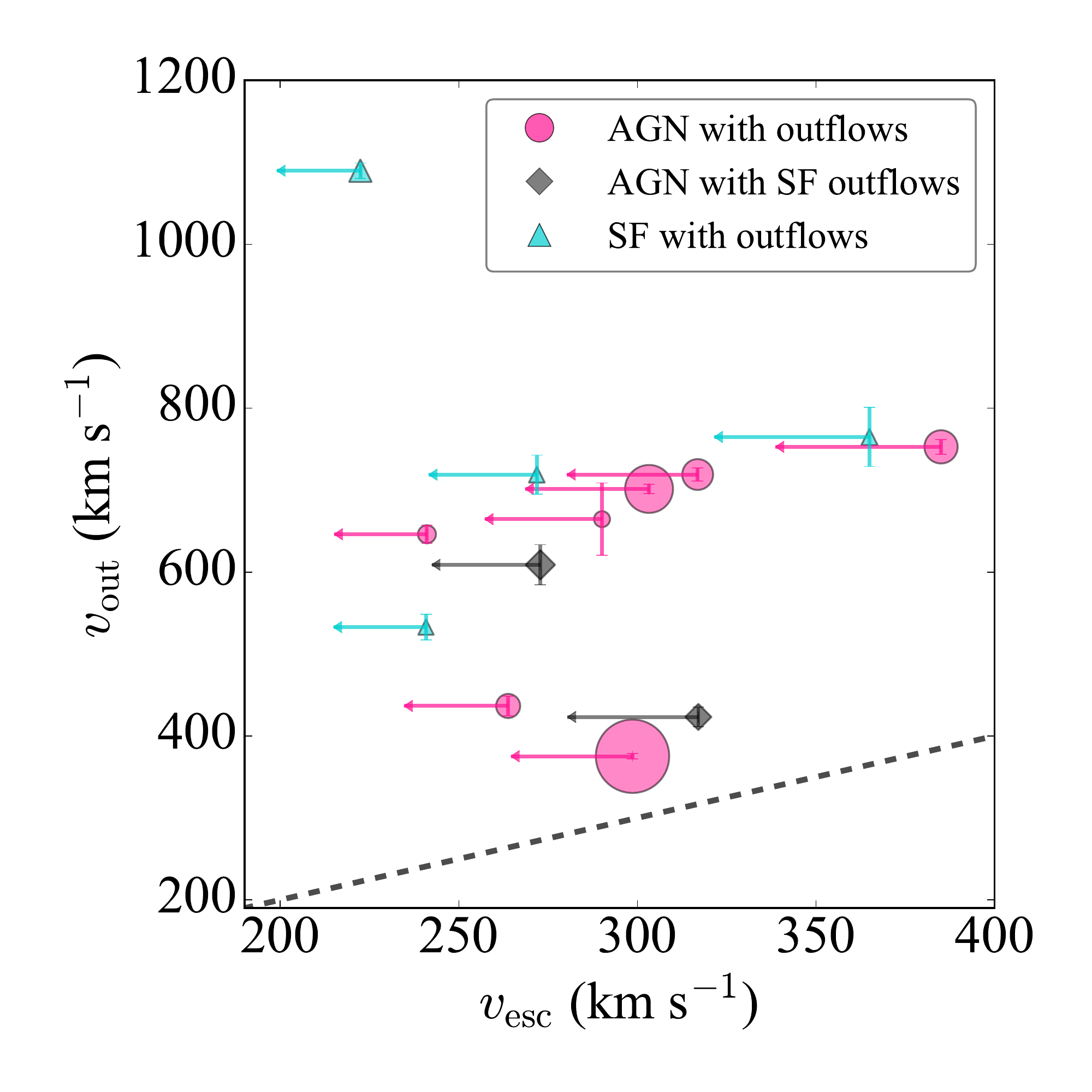}
\caption{Relation between the velocity measured for the outflows and the modeled escape velocity assuming galaxies live in massive dark matter halos consistent with the cosmological picture of $\Lambda$CDM. In all cases, the wind speed is comparable or above that needed to escape the dark matter halo. To guide the eye, the dashed line indicates a 1-to-1 relation.}
\label{fig:vout}
\end{figure}
%

In order to test whether these winds are capable of escaping the gravitational potential of their host halos, we calculated the escape velocity of each galaxy.  We assume an NFW dark matter density profile \citep{Navarro1996} and use abundance matching \citep{Moster2013} to estimate the halo mass from stellar mass for each galaxy.  The escape velocity is 
\begin{equation}
v_{\rm esc}(r)^2 = 2\;|\Phi(r)|
\end{equation}
where $\Phi$ is the gravitational potential corresponding to a spherical NFW profile \citep{Lokas2001}.
The escape velocities for all 13 outflow galaxies were calculated at $r = 0$ kpc.  
Calculating escape velocity from the center of a cuspy dark matter halo constitutes an upper limit, as escape velocity decreases with radius.  A cored dark matter profile for dwarfs, as has been suggested from observations of some dwarf galaxies \citep{Adams2014, Oman2015}, will also result on a lower escape speed, facilitating the gas removal.  Halo mass predictions and escape velocities for each outflow galaxy are listed in Table~\ref{table:outflow}.

\begin{center}
\begin{figure}
\includegraphics[width=\linewidth,clip]{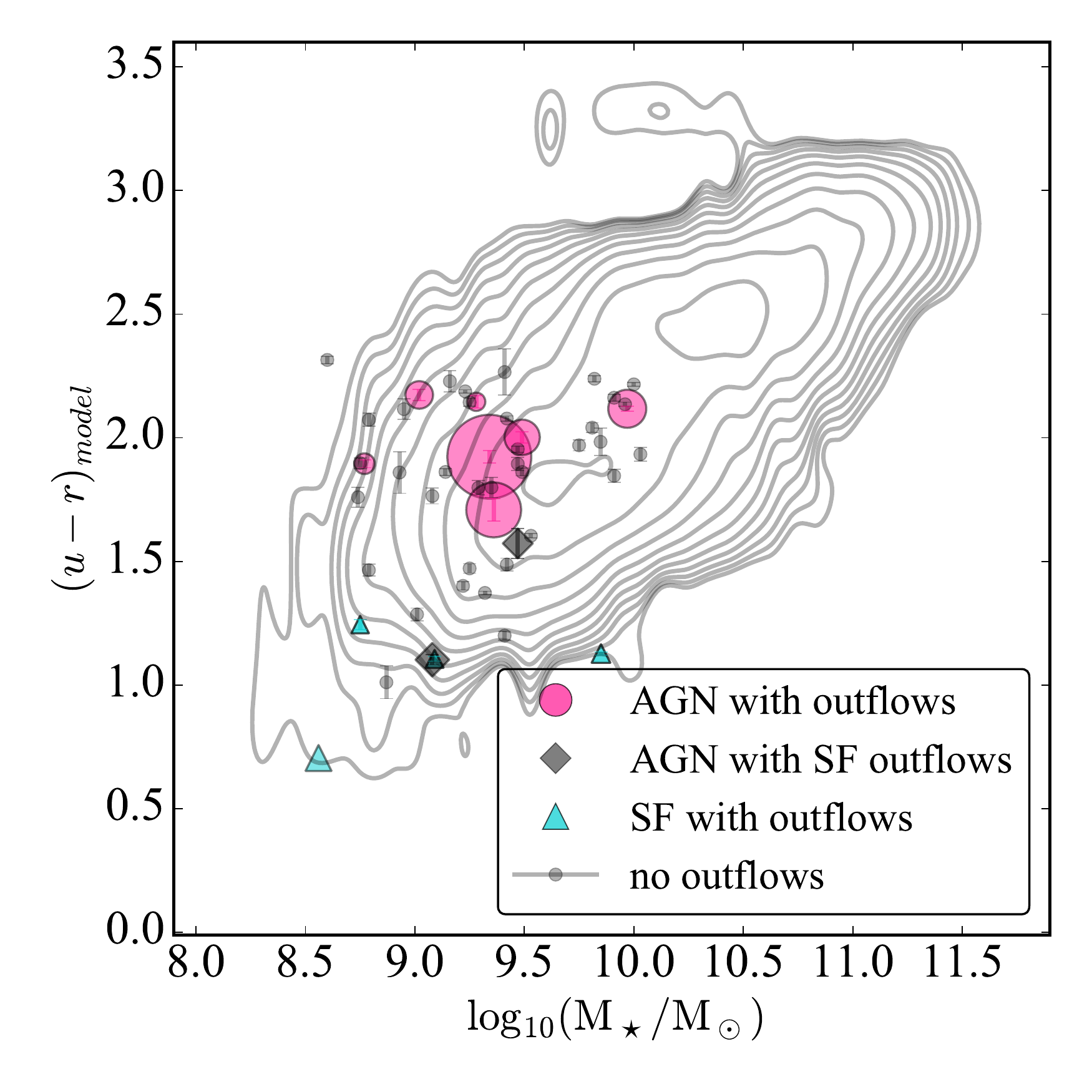}
\caption{The $u-r$ model magnitude colors from SDSS DR 8 are plotted against the MPA-JHU stellar mass.  The colors are corrected for galactic extinction, following \cite{Schlegel1998} and the contours are from \cite{schawinski2014}.  The photometry for NGC 1569 were measured in GALEX \cite{galex2017} and converted to SDSS $ugriz$ via the Python code pysynphot.  The outflow symbol sizes are scaled to the fraction of [\ion{O}{3}] flux contained in the outflows.}
\label{fig:color}
\end{figure}
\end{center}

Figure~\ref{fig:vout} compares the velocity of the outflow $v_{\rm{out}}$ to an upper limit of the escape velocity $v_{\rm{esc}}$.  As discussed in Section~\ref{sec:kinematics}, we measure $v_{\rm out}$ in our targets by taking the velocity blueshift corresponding to the $80\%$ width of the detected broad lines plus the offset from the system's velocity (Fig.~\ref{fig:OIII_fit}).  The fit is performed on spectra extracted from within $R_{50}$.  The length of the horizontal arrows in Fig.~\ref{fig:vout} shows the impact of changing the assumption of halo concentration from $c=15$ to $c=8$.  The symbol sizes in Figs~\ref{fig:vout} and \ref{fig:color} are proportional to the percentage of [\ion{O}{3}] flux in the outflow (i.e. the ratio of the flux of the broad component to the total [\ion{O}{3}] line flux).  The velocities plotted in Fig.~\ref{fig:vout} as well as the flux ratios that determine the symbol sizes are listed in Table~\ref{table:outflow} and the details of their calculations can be found in Section~\ref{sec:kinematics}.

In all cases the outflow velocities surpass $v_{\rm esc}$, suggesting that the ionized gas entrained in the outflow will become unbound from the galaxy and, later, its dark matter halo.

%

\subsection{Feedback}
\label{sec:feedback}
%

AGN feedback has been studied extensively at higher masses, invoking winds to explain both suppressing \citep{DiMatteo2005, Fabian2012}) and enhancing \citep{DeYoung1989, Gaibler2012, Ishibashi2013} effects on star formation.  Studies of galaxy evolution placing AGN on the color-mass diagram imply a scenario where AGN mark a sudden transition from blue, star forming galaxies into quiescent red galaxies \citep{MartinChristopher2007}.  A large body of theoretical \citep{Khalatyan2008, Trayford2016,Taylor2015} and observational \citep{Salim_2007, Nandra2007, Donoso2012} work has been conducted in the high mass regime to investigate whether AGN can regulate star formation by means of expelling gas or by disrupting cooling flows that would otherwise fall in and fuel star formation.

Figure~\ref{fig:color} shows that dwarfs with detected outflows associated with AGN (pink circles) have intermediate colors that strike between the blue, actively star forming population, and the redder, quiescent population.  Their position is at least qualitatively consistent with the evolutionary scenario proposed at higher masses.  Non-AGN star forming galaxies with outflows (cyan triangles) are among the bluest in our sample, as is expected from very young and active stellar populations.  In the bluest AGN hosts discussed here, the outflows appear to be at least partially powered by coincident stellar processes (gray diamonds).  These galaxies retain their blue colors despite having fast outflows.  This could imply that the outflows are not quenching the star formation, or it could be that the timescale for galaxy color evolution is longer than the specific stage of the outflows at which we are capturing these objects.

In a scenario where AGN-driven outflows suppress star formation, one might expect to find such outflows in redder galaxies.  An alternative explanation for this trend is that optically selected AGN are more easily detected in galaxies with lower star formation rates.  Though our selection criteria relies exclusively on emission line properties and does not consider photometry at all, 
contribution to gas ionization from young stellar populations can obscure optical emission line signatures of AGN.  Our star forming galaxies lack all available indicators of AGN; however, we cannot definitively rule out the presence of faint AGN activity coincident with active star formation in our control sample.

All of the outflow galaxies discussed here have H$\alpha$ equivalent widths $(\rm{EW}_{H\alpha})$ well above $3\,\rm$\mbox{\AA}, and thus cannot be considered retired $(\rm{EW}_{H\alpha} < 3\,$\mbox{\AA}$)$ or quenched  $(\rm{EW}_{H\alpha} < 0.5\,$\mbox{\AA}$)$ \citep{CidFernandes2011}. If the presence of outflows does imply removal of all of the gas in the dwarf galaxy, we are likely catching these galaxies in a phase prior to quenching.  This is in contrast to other observational work suggesting AGN feedback works to suppress star formation after an initial quenching event, likely due to tidal stripping by a neighboring galaxy \citep{Penny2018}.  
Almost all of the galaxies in our sample are isolated (see Appendix~\ref{sec:environment}), so internal processes likely govern star formation in these objects.

AGN-driven outflows are suspected to play a major role in quenching star formation in the most massive galaxies.  The AGN-driven outflow velocities we report in this work are comparable to those measured in some higher mass galaxies (e.g., \citet{Harrison2014, Rupke2017, Baron2018}).  If AGN-powered winds are suspected to quench star formation in those high mass galaxies, the similar wind velocities that we measure could be a plausible quenching mechanism in the dwarf regime as well.  The ability of these AGN to quench their host galaxies ultimately depends on the fraction of gas mass involved in the outflow.  Due to substantial slit losses and unconstrained electron densities, estimates of outflow gas mass using our current data would span many orders of magnitude.  Follow-up observations of two of the AGN presented here and a thorough discussion of mass, kinetic energy, and ionization conditions of the outflows will be presented in a follow up paper (W. Liu et al., in prep.)

\section{Summary}
\label{sec:summary}

We have presented the detection and kinematic measurements of extended outflows in isolated dwarf galaxies.  Our Keck LRIS longslit data provided spatially resolved spectroscopy, in all cases revealing galaxy-wide winds with velocities exceeding the escape velocities of their dark matter halos.  Based on emission line ratios of the outflow components, we found the central AGN to be the dominant driving mechanism in at least six of 13 galaxies.  We summarize our conclusions below.

\begin{enumerate}
    \item Thirteen of 50 dwarf galaxies both with and without optical signs of AGN show galaxy-wide winds.    Nine of these 13 galaxies are classified as AGN according to their narrow line flux ratios.   The remaining four have no optical or IR signatures of AGN activity and are thus classified as star forming.
    
    \item  We were able to measure BPT emission line ratios for the broad components of 12 of the 13 galaxies with outflows.  The four galaxies classified as SF based on their narrow line ratios have outflow line ratios consistent with SF ionization plus some contribution from shocks.  
    
    Of the nine AGN, two have outflow line ratios that fall in the BPT composite region, indicating that their outflows might be at least partially powered by star formation.  One AGN had insufficient outflow flux to place it on the BPT diagram.  The remaining six AGN have outflow line ratios that exceed the theoretical maximum ionization possible with star formation alone.  Therefore, we report the detection of AGN driven outflows in at least six dwarf galaxies.
    
    
    
    \item Outflow velocities were measured to be $375 - 1090 \,\rm{km\, s}^{-1}$ for galaxies with and without AGN, ranging in stellar mass $\sim 4\times10^8 - 9\times10^9 M_\odot$.  The outflow velocities in all 13 galaxies are sufficient to escape their dark matter halos.
    
    \item SF galaxies and AGN show differences in the line profiles of their outflow components.  AGN outflow profiles tend to be more blueshifted and slightly narrower than those of SF galaxies. Outflows with BPT composite line ratios have line profiles similar to those in SF galaxies.   We speculate that the differences may be due to the differing physical placement of the wind sources for each of these type of objects.
    
    \item AGN-driven outflows tend to carry a larger fraction ($5-50\%$) of the total amount of [\ion{O}{3}] flux than the SF outflows ($4-8\%$).  The outflows carrying the largest fractions of ionized gas tend to populate redder galaxies.  The placement of the AGN-driven outflows on the color-mass diagram is suggestive of ongoing star formation suppression due to the influence of the AGN.  
\end{enumerate}

The outflow velocities reported here serve as the first directly observed AGN-driven outflows in dwarf galaxies and offer vital observational constraints necessary to extend realistic feedback models into the low mass regime.  Detailed observational followup is needed to constrain the gas mass, spatial extent, and energetics of the outflows in these galaxies.  Careful consideration of the potential contribution of shock heating to line ratios is also necessary.

Fast outflows exist in a third of the low-mass AGN hosts in our sample of 50 dwarfs, so a systematic search for AGN-driven outflows in a much larger parent sample of dwarf galaxies is warranted.  

\vspace{1cm}

{\bf Acknowledgements}\\
We thank the anonymous referee, whose careful reading and constructive suggestions helped improve and clarify this manuscript. 
Support for this program was provided
by the National Science Foundation, under
grant number AST 1817233.  Additional support
was provided by NASA through a grant from the
Space Telescope Science Institute (Program AR-
14582.001-A), which is operated by the Association of
Universities for Research in Astronomy, Incorporated,
under NASA contract NAS5-26555.  LVS acknowledges support 
from the Hellman Foundation.
The data presented herein were obtained at the W. M.
Keck Observatory, which is operated as a scientific partnership 
among the California Institute of Technology,
the University of California and the National Aeronautics 
and Space Administration. The Observatory was made 
possible by the generous financial support of the
W. M. Keck Foundation.
The authors wish to recognize and acknowledge the
very significant cultural role and reverence that the 
summit of Mauna Kea has always had within the indigenous
Hawaiian community. We are most fortunate to have the
opportunity to conduct observations from this mountain.
Some of the data presented herein were obtained using the UCI
Remote Observing Facility, made possible by a generous gift 
from John and Ruth Ann Evans. 
This research has made use of the NASA/IPAC Extragalactic Database (NED), which is operated by the Jet Propulsion Laboratory, California Institute of Technology, under contract with the National Aeronautics and Space Administration.


\begin{longtable*}[t]{| l l l l r r r r c c c|}
\hline
Name & $z$& $M_*$ & $M_{\rm halo}$  & $v_0$ & $W_{80}$  &$v_{\rm out}$ & $v_{\rm esc}$ &$F_{\rm b}/F_{\rm tot}$ & narrow & broad \\ 
(1) & (2) & (3)  & (4) & (5) & (6) & (7) & (8) & (9) & (10) &(11)\\\hline
\textbf{AGN}&&&&&&&&&& \\ 
\hline
J081145.29+232825.7 & $0.016$ &$9.02$  &$11.02$  &$-120\pm11 $ & $634 \pm12  $& $  437\pm167 $ & $264\pm29$  &$0.1002$& AGN  & AGN \\
J084025.54+181858.9 & $0.015$ &$9.28$  &$11.13$  &$ 149\pm44 $ & $1627\pm44  $& $  665\pm62  $ & $290\pm33$  &$0.0467$& AGN  & --- \\
J084234.51+031930.7 & $0.028$ &$9.34$  &$11.17$  &$ -72\pm2.5$ & $607 \pm3.5 $& $  375\pm4.3 $ & $299\pm34$  &$0.5060$& Comp & AGN \\
J090613.75+561015.5 & $0.045$ &$9.36$  &$11.19$  &$-128\pm4.4$ & $1147\pm5.8 $& $  701\pm7.2 $ & $303\pm35$  &$0.3046$& AGN  & AGN \\
J095418.16+471725.1 & $0.032$ &$9.49$  &$11.24$  &$ -79\pm5.4$ & $1280\pm7.9 $& $  719\pm9.6 $ & $317\pm37$  &$0.1542$& AGN  & AGN \\
J100551.19+125740.6 & $0.0093$&$9.97$  &$11.49$  &$-176\pm7.6$ & $1154\pm8.8 $& $  753\pm17  $ & $385\pm46$  &$0.1740$& AGN  & AGN \\
J100935.66+265648.9 & $0.014$ &$8.77$  &$10.90$  &$-119\pm9.8$ & $1055\pm11  $& $  646\pm15  $ & $241\pm26$  &$0.0587$& AGN  & AGN \\
\hline
\textbf{AGN w/SF outflows }&&&&&&&&&& \\ 
\hline
J010005.94$-$011059.0& $0.050$ &$9.47$  &$11.24$  &$  41\pm11 $ & $929 \pm12  $& $  423\pm16  $ & $317\pm37$  &$0.0616$ & Comp & Comp \\
J144252.78+205451.6  & $0.043$ &$9.08$  &$11.06$  &$  -8\pm23 $ & $1201\pm25  $& $  609\pm34  $ & $273\pm30$  &$0.0765$ & AGN & Comp\\
\hline
\textbf{Star Forming}&&&&&&&&&&\\
\hline
J101440.21+192448.9  & $0.028$  &$8.75$  &$10.90$  &$  24\pm14 $ & $1114\pm16  $& $  533\pm21  $ & $241\pm26$  &$0.0431$& SF & SF \\
J130724.63+523715.2  & $0.026$  &$9.09$  &$11.05$  &$   2\pm22 $ & $1442\pm24  $& $  719\pm33  $ & $272\pm30$  &$0.0452$& SF & SF \\
J171759.66+332003.8  & $0.015$  &$9.85$  &$11.42$  &$ -11\pm35 $ & $1509\pm36  $& $  765\pm50  $ & $365\pm44$  &$0.0482$& SF & SF \\
NGC 1569             & 2 Mpc    &$8.56$  &$10.81$  &$  23\pm8.8$ & $2225\pm9.4 $& $ 1090\pm13  $ & $222\pm23$  &$0.0878$& SF & SF \\
\hline
\caption{
Details of fits to spectra extracted within $R_{50}$ of each galaxy, unless otherwise specified.\\
(1) Full SDSS names of all 13 galaxies with extended outflows \\
(2) Redshift\\
(3) Stellar mass reported in the MPA-JHU catalog\\
(4) Halo mass determined via abundance matching using the method described in the text and the stellar mass listed in (2)\\
(5) Mean velocity of the outflow, in km s$^{-1}$, relative to the center of the narrow component (negative$=$blueshifted)\\
(6) Width containing 80\% of the flux of the outflow component, in km s$^{-1}$.\\
(7) Outflow velocity defined as $v_{\rm out} = -v_0 + \frac{W_{80}}{2}$, in km s$^{-1}$\\
(8) Velocity required to escape the dark matter halo (3) with an NFW profile from $r=0$, in km s$^{-1}$\\
(9) Fraction of [\ion{O}{3}] flux in the outflow, calculated as the ratio of the flux in the broad component, $F_{\rm b}$, to the total flux in broad $+$ narrow components, $F_{\rm tot}$\\
(10,11) Decomposed BPT classifications for the bound and outflowing gas.  In order to optimize S/N, these fits were performed on spectra extracted from customized apertures for each galaxy.
}
\label{table:outflow}
\end{longtable*}

\appendix

\section{Black Hole Masses}
\label{sec:M_bh}
%
Line broadening from the Broad Line Region (BLR) associated with the AGN is sometimes observed in Balmer emission lines (see Figure~\ref{fig:fits}) and can be used to measure BH masses.  Due to the small spatial extent of the BLR and the low transition probability of forbidden lines, BLR emission does not contribute to the line profiles of [\ion{O}{3}]$\lambda$5007, which we use to trace the kinematics of fast-moving winds throughout the galaxy.  In this section, we explore the presence of an \textit{additional} BLR component in H$\alpha$ (purple line in the third panel of Fig.~\ref{fig:fits}), associated with the BH itself, not to be confused with the broadened components associated with galaxy-scale outflows (orange lines in Fig.~\ref{fig:fits}).

We detect a previously unreported broad H$\alpha$ line in J084234.51+031930.7 and measure its virial mass to be $M_{BH} = 6.92\times10^5\, M_\odot$, using RGG13 
Equation 5, which relies on the $R_{BLR} - L$ relation of \cite{bentz2013}. 
\begin{equation}
\begin{split}
    \log(M_{BH}/M_\odot) = \log(\epsilon) + 6.57 
    + 0.47\,\log\left(\frac{L_{\rm{H}\alpha}}{10^{42}\rm{\,erg\,s}^{-1}}\right) 
    + 2.06\,\log\left(\frac{\rm{FWHM}_{\rm{H}\alpha}}{10^3\rm{\,km\,s}^{-1}}\right)
\end{split}
\end{equation}

Black hole masses for 5 more AGN in this sample are reported in the literature using this approach, yet once outflow components are included in the [\ion{N}{2}], H$\alpha$ narrow line model, we find no additional BLR H$\alpha$ components in the other  AGN discussed here.  We therefore caution that line broadening due to outflowing gas could masquerade as a BLR broad line, so it is necessary to check for outflow contributions in forbidden lines such as [\ion{O}{3}] and model them whenever possible when deriving black hole masses from heavily blended complexes such as H$\alpha\,+\,$[\ion{N}{2}].

\section{Spatial Properties of Outflows}
\label{apx:spatial}

Figure 11 shows the spatial properties of the outflows of the remaining 12 galaxies, as in Fig.~\ref{fig:spatial}.

\begin{center}
\label{fig:appendix_spatial}
\begin{figure}[b]
\includegraphics[width=0.31 \linewidth,clip]{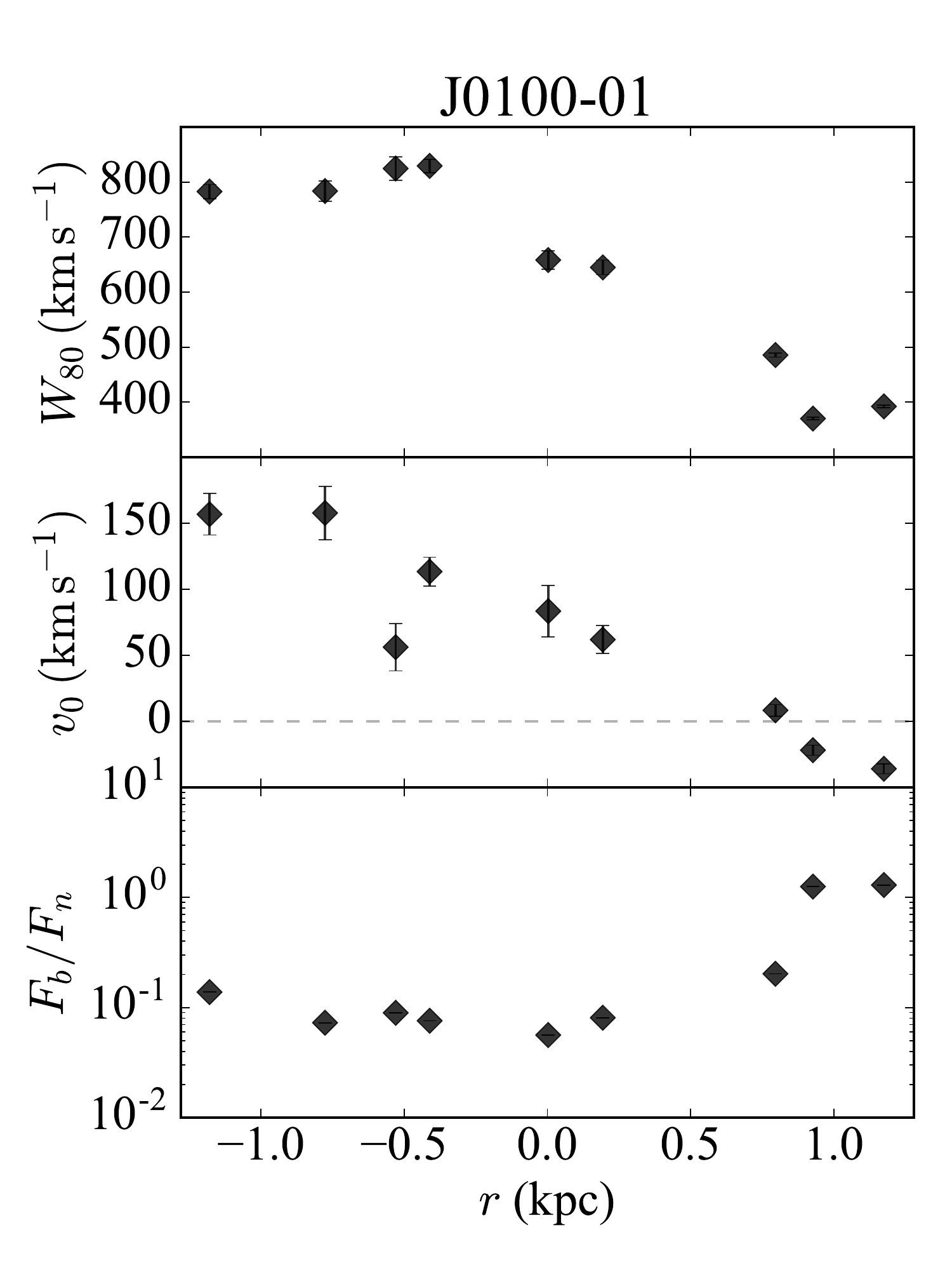}
\includegraphics[width=0.31 \linewidth,clip]{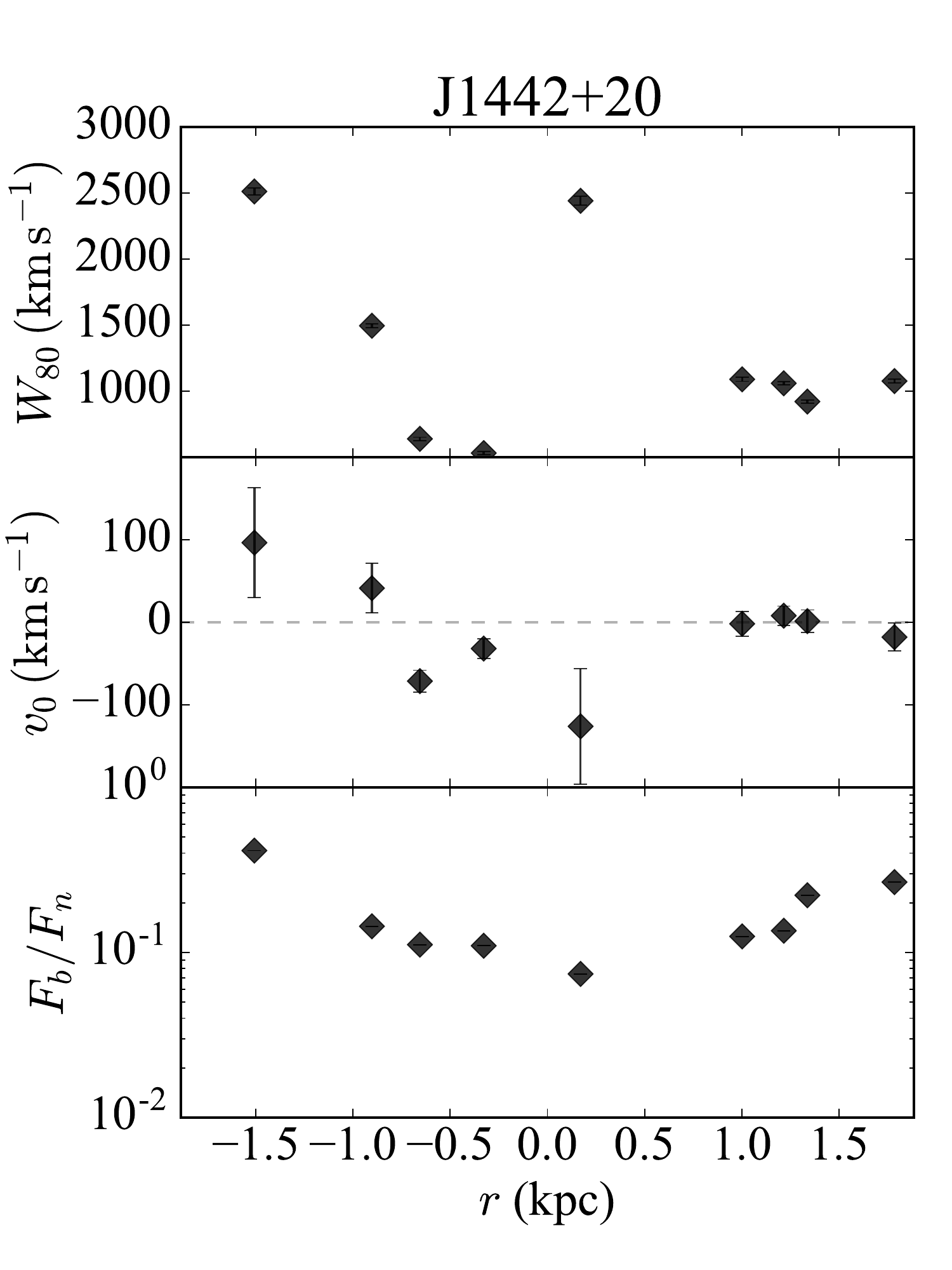}
\includegraphics[width=0.31 \linewidth,clip]{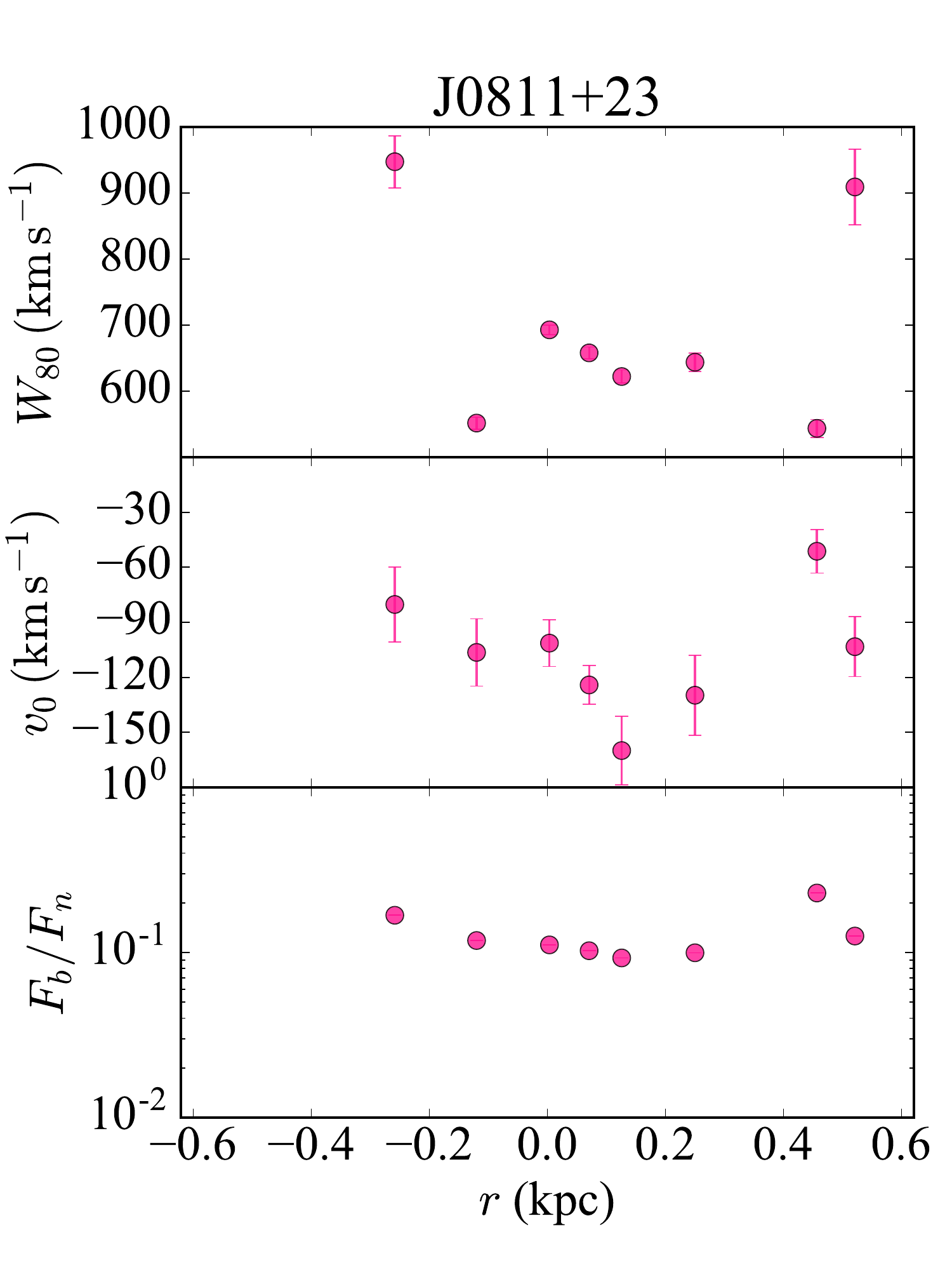}\\
\includegraphics[width=0.31 \linewidth,clip]{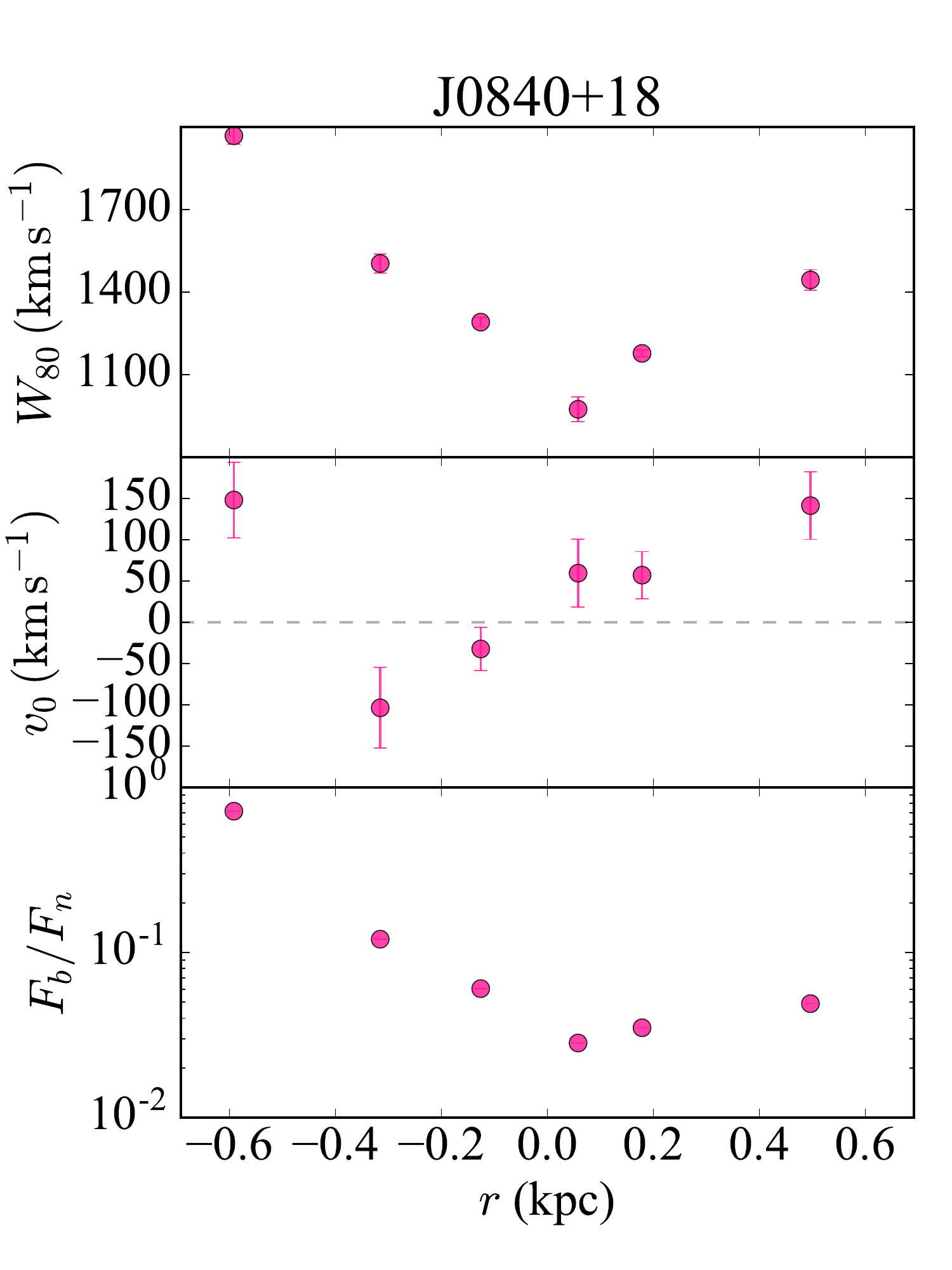}
\includegraphics[width=0.31 \linewidth,clip]{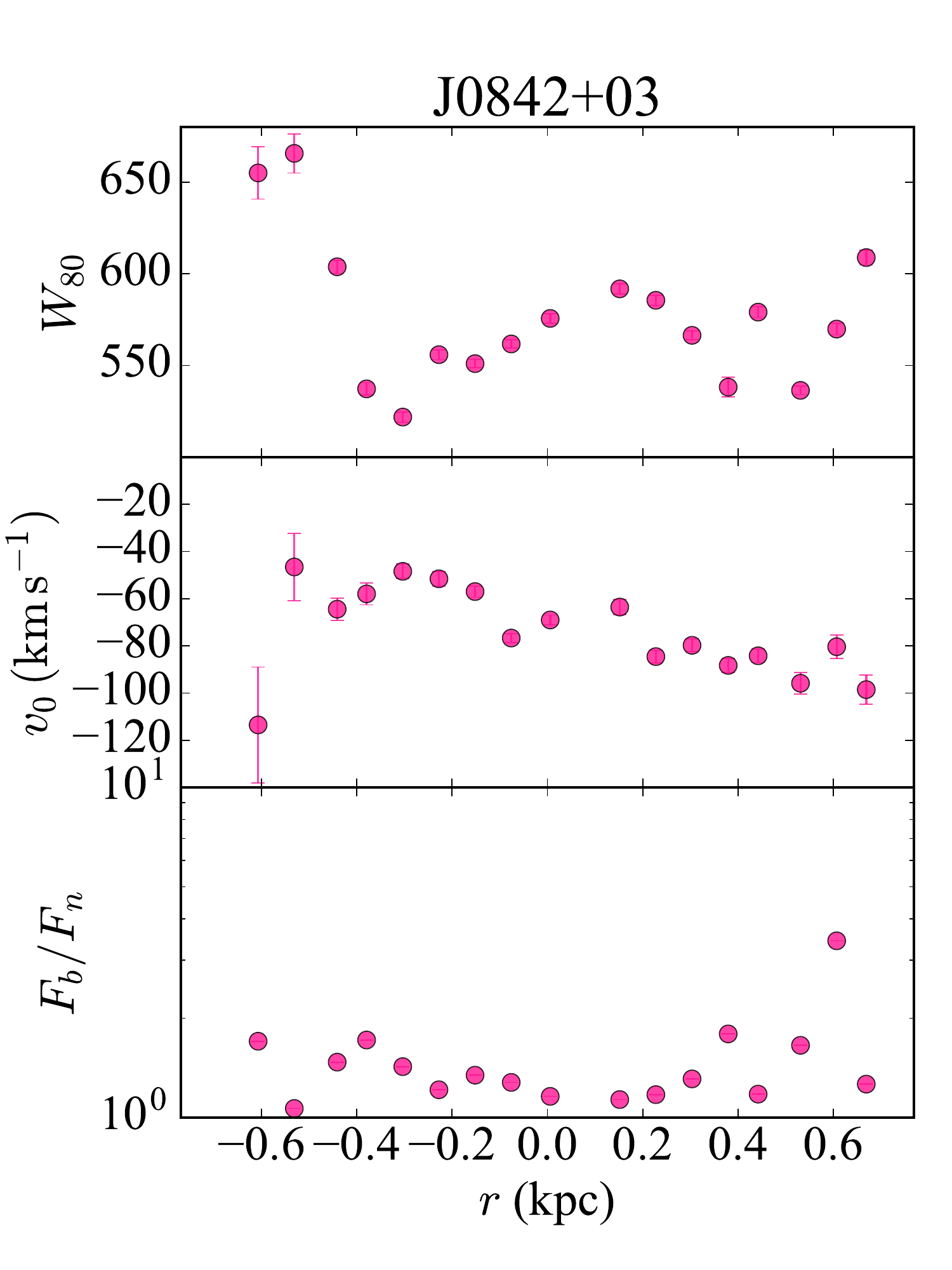}
\includegraphics[width=0.31 \linewidth,clip]{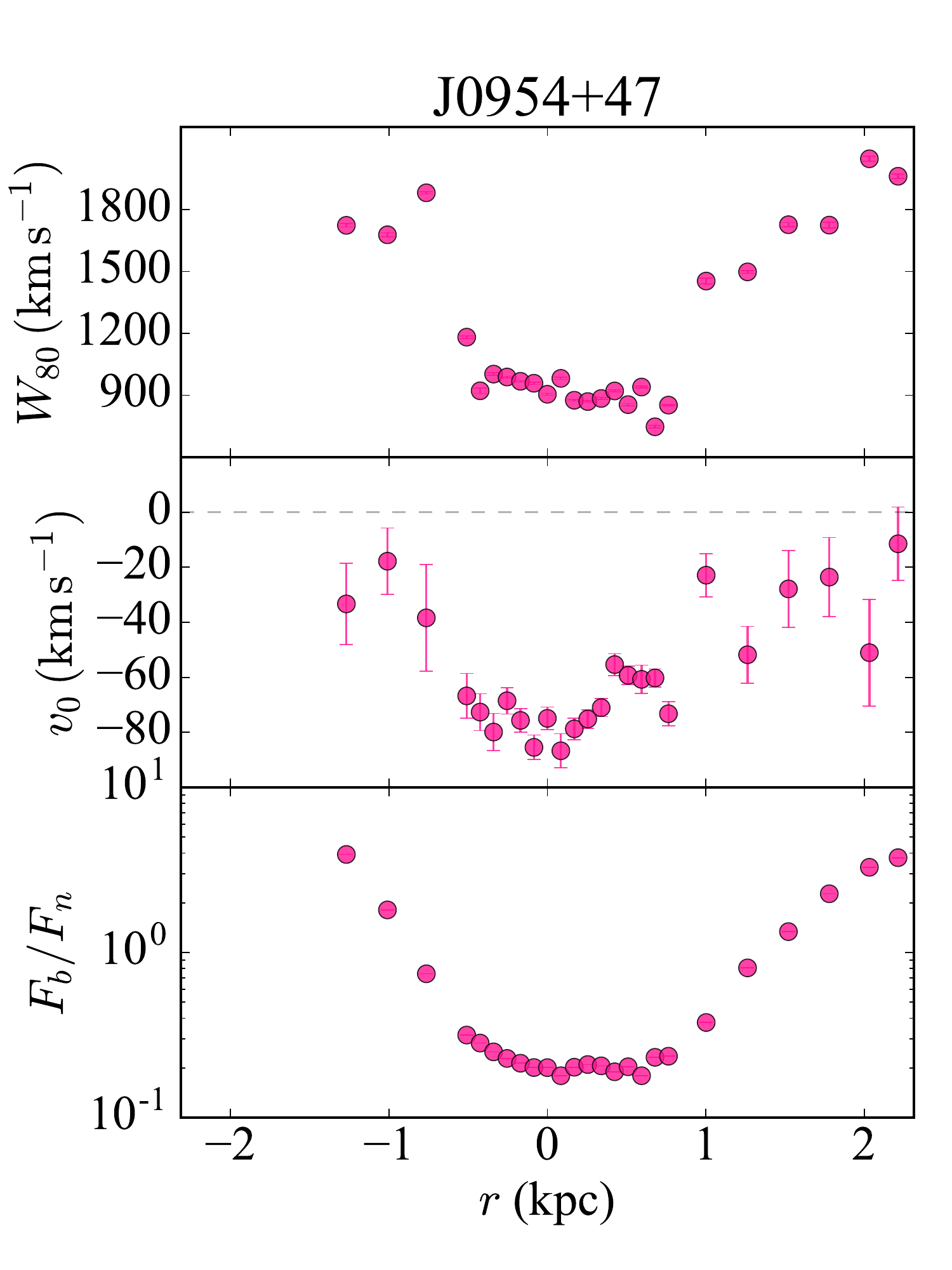}\\
\includegraphics[width=0.31 \linewidth,clip]{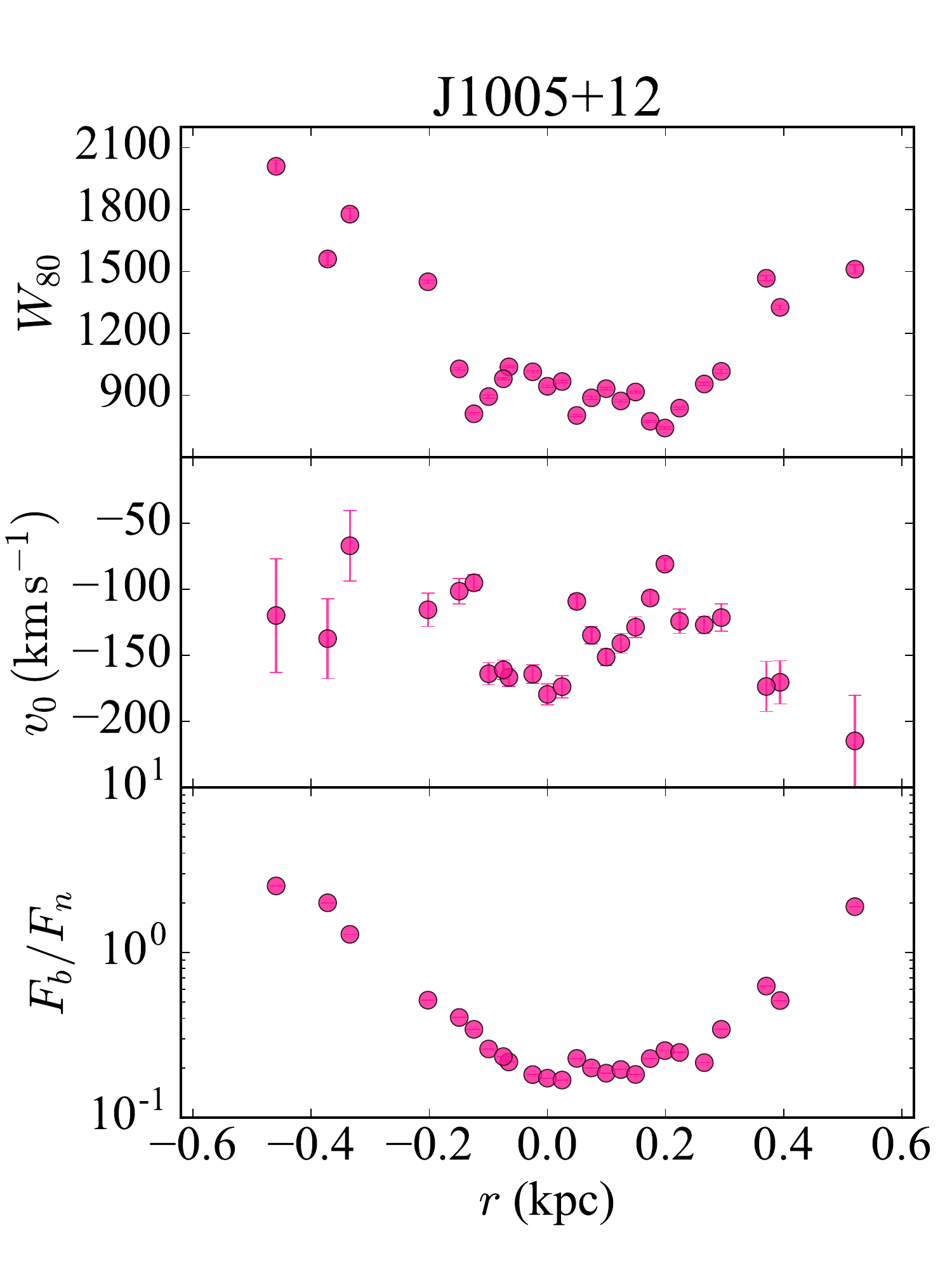}
\includegraphics[width=0.31 \linewidth,clip]{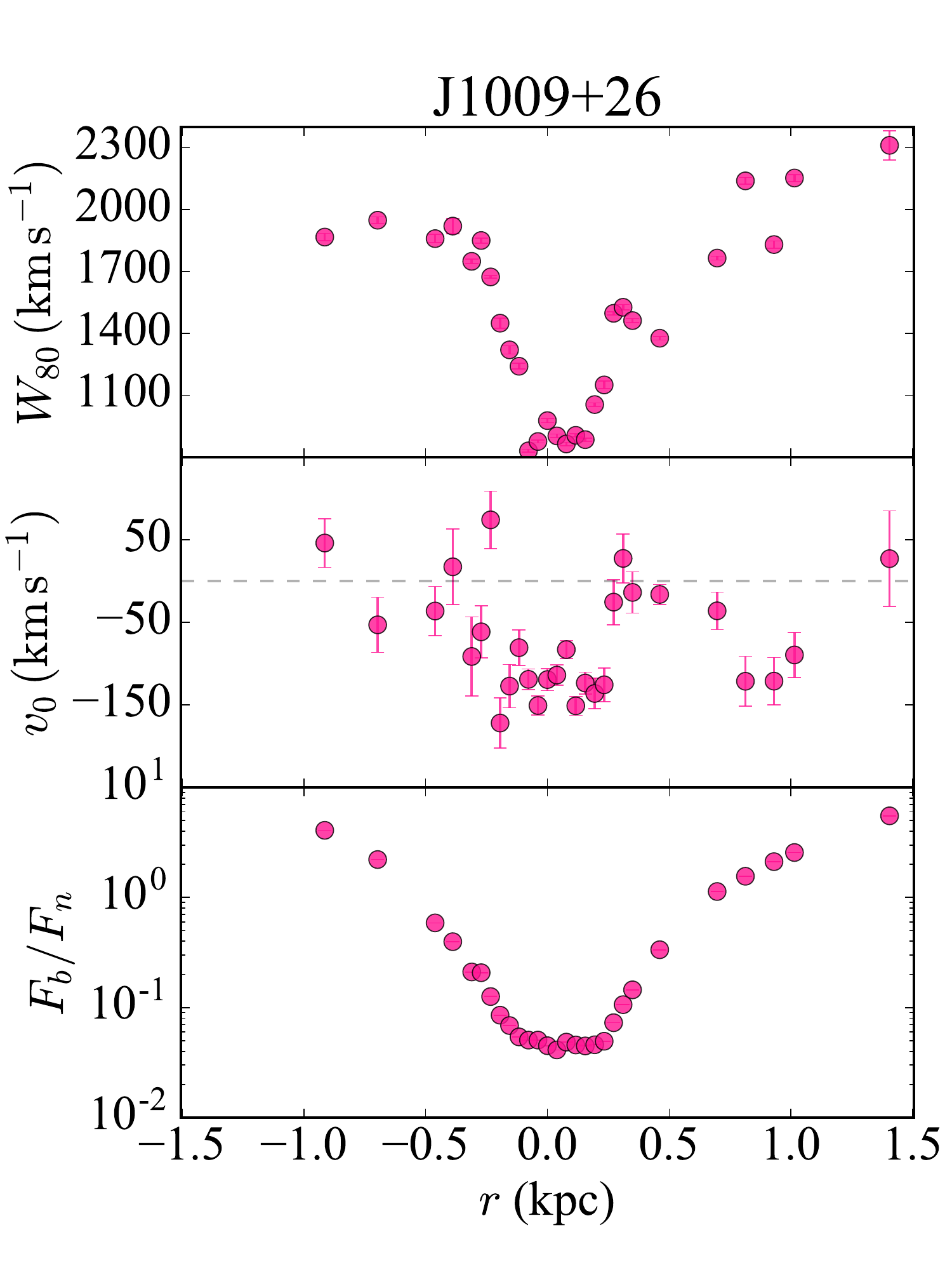}
\includegraphics[width=0.31 \linewidth,clip]{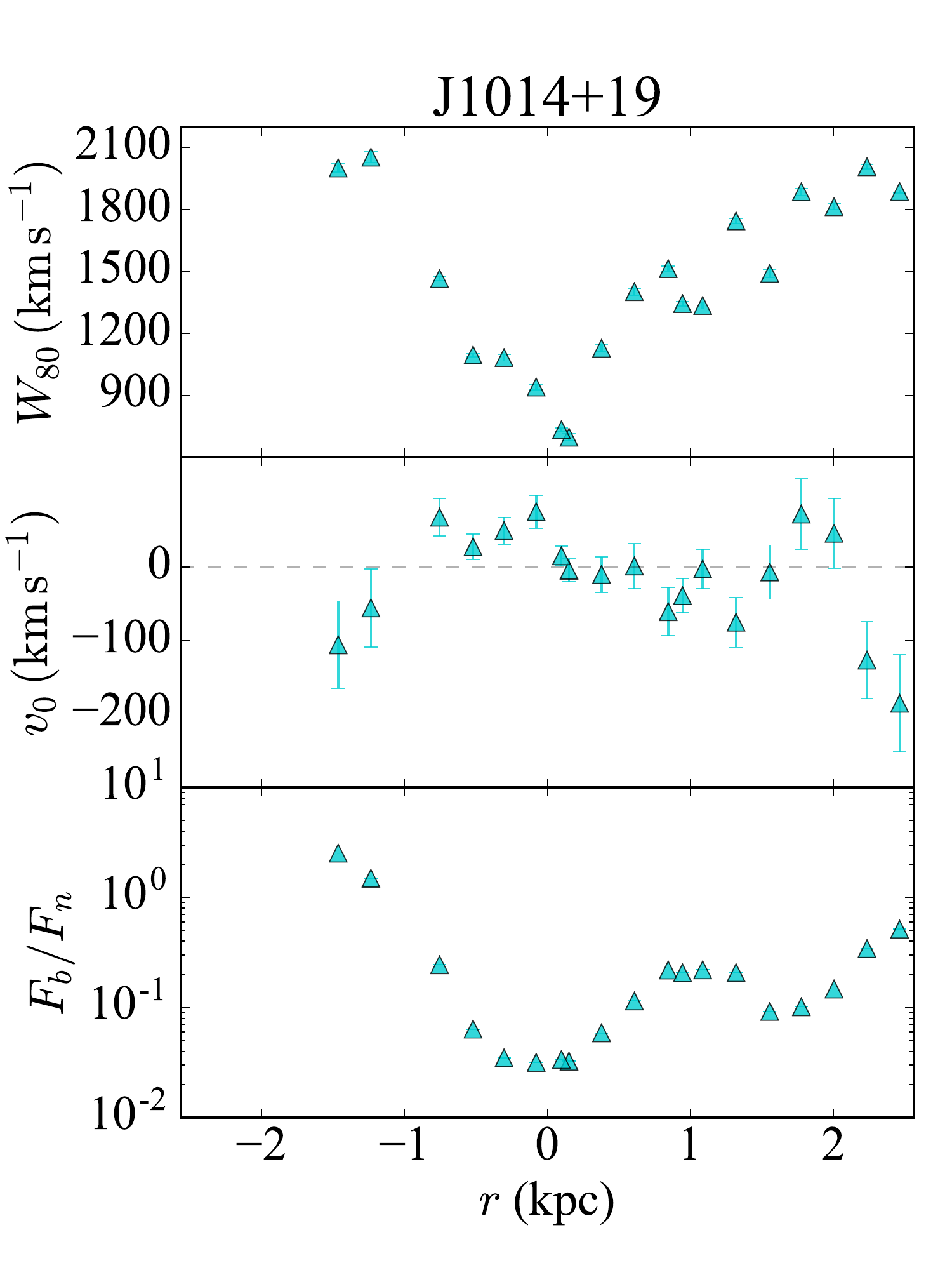} \\
\end{figure}
\end{center}
\begin{center}
\begin{figure}
\includegraphics[width=0.31 \linewidth,clip]{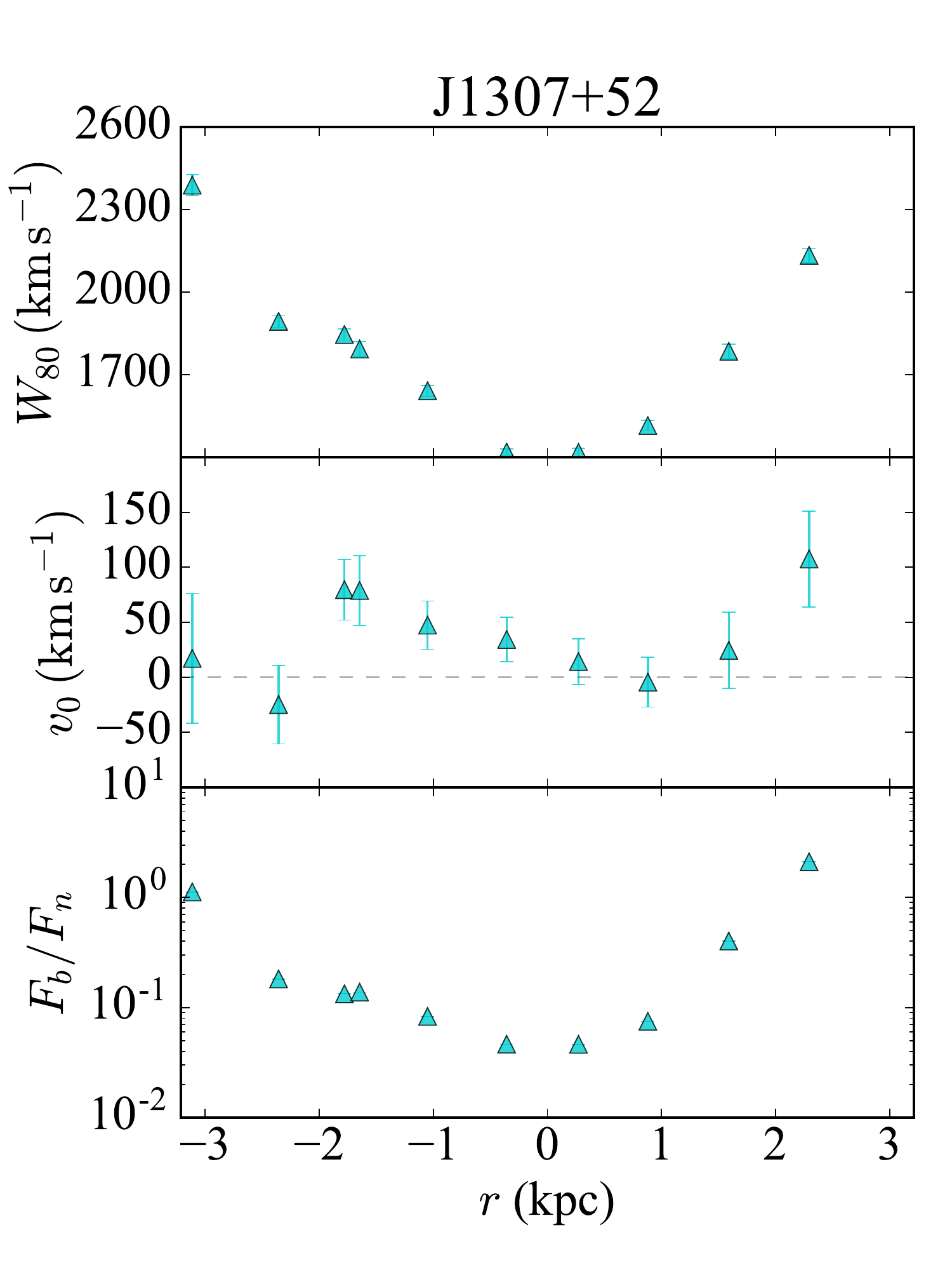}
\includegraphics[width=0.31 \linewidth,clip]{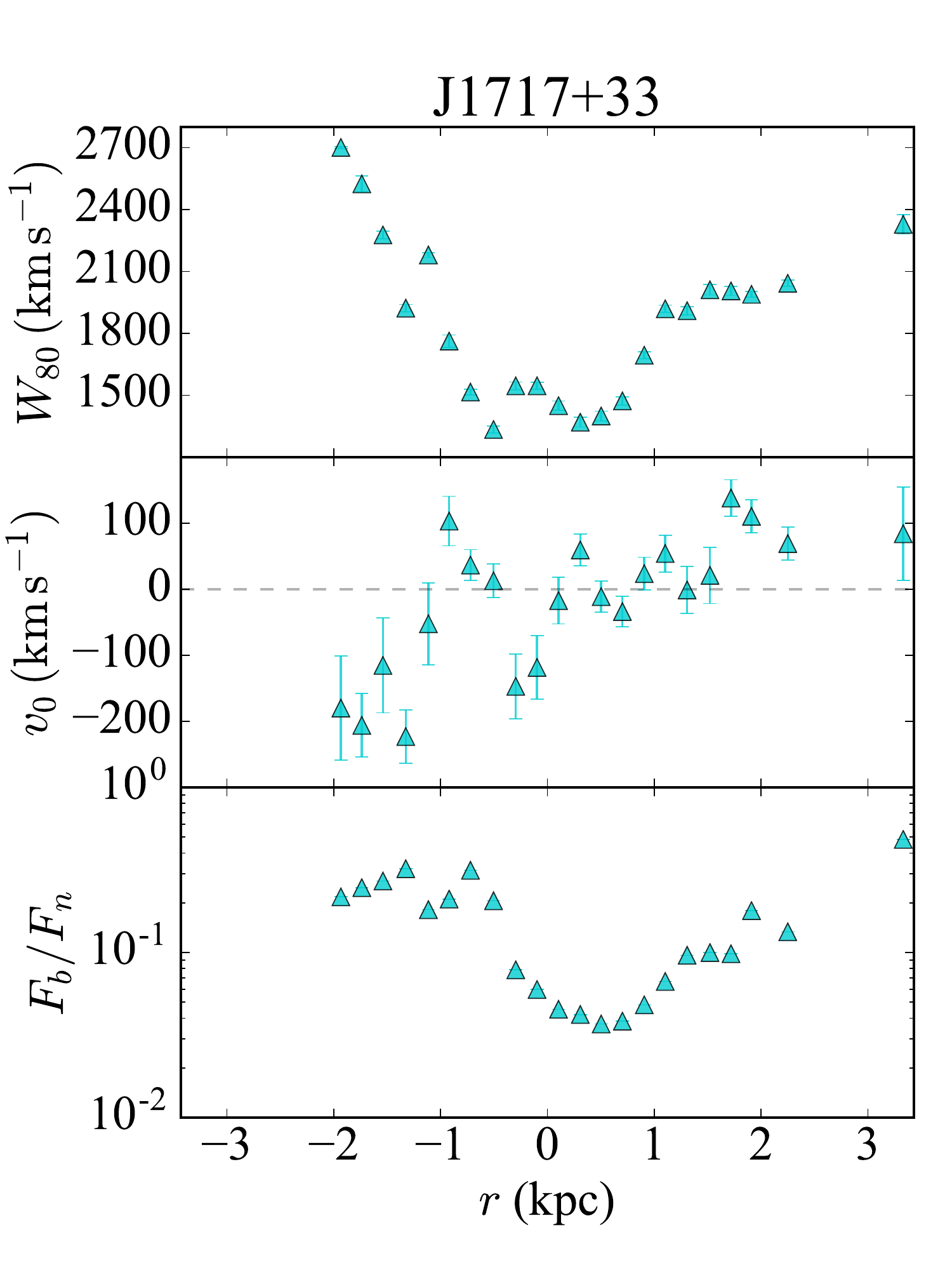}
\includegraphics[width=0.31 \linewidth,clip]{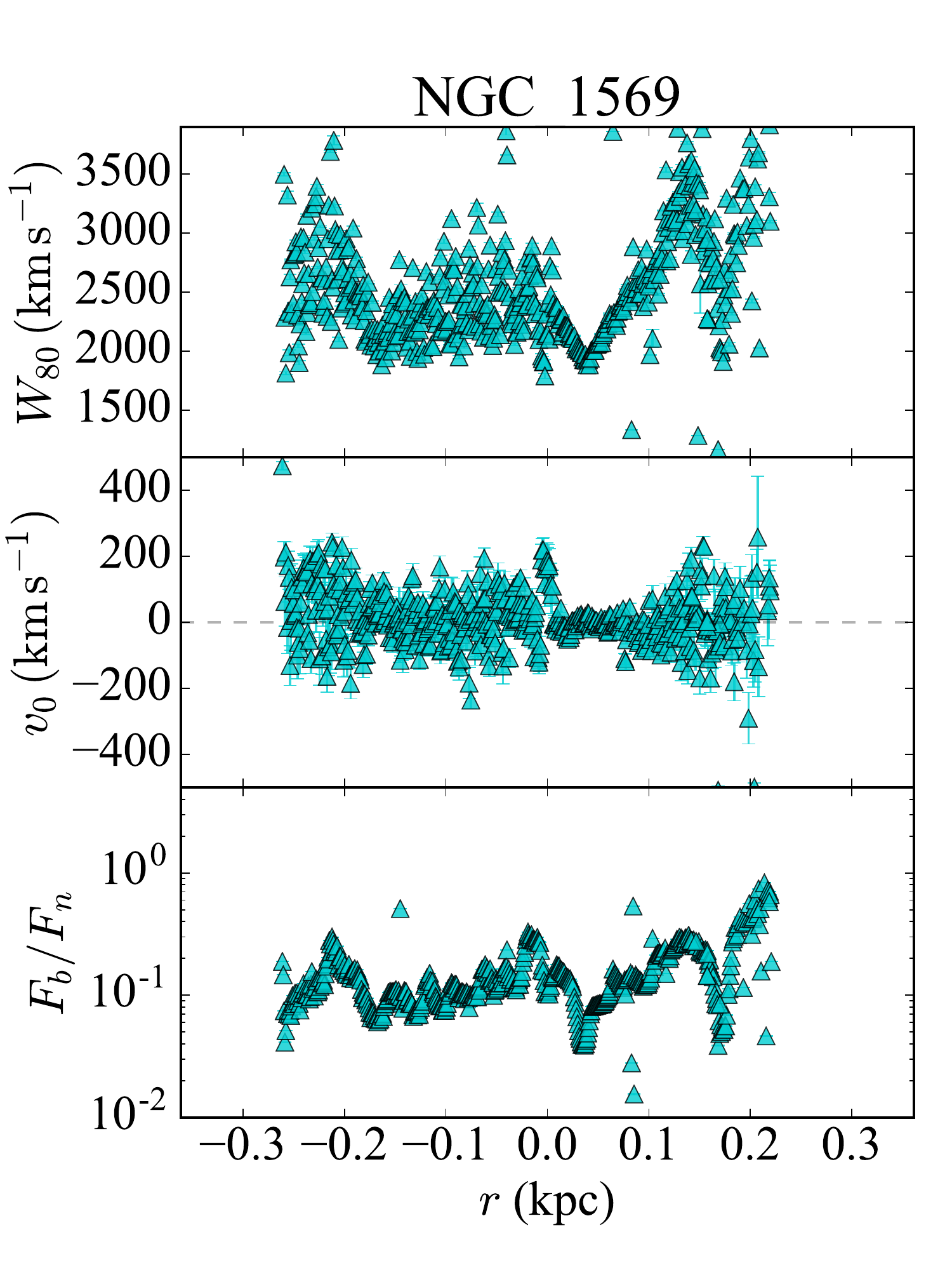}
\caption{Spatially resolved properties of outflows are presented for the remaining 12 galaxies, as in Fig.~\ref{fig:spatial}.  Consistent with the scheme used throughout this paper, AGN-driven outflows are plotted with pink circles, star forming with cyan triangles, and AGN with SF outflows with black diamonds.}
\end{figure}
\end{center}

\vspace{-1.5 cm}
\section{Environment}
\label{sec:environment}
Though environment was not considered when building our sample, almost all of our galaxies are isolated.  A query of all SDSS DR 14 objects with spectra within 120 arcminutes of each galaxy reveals that none of the galaxies are within a comoving distance of 1 Mpc of a more massive companion.  This query was followed up on the NASA/IPAC Extragalactic Database in order to check for objects not covered by SDSS. J171759.66+332003.8 is 370 kpc from a galaxy of comparable size ($\log{M_*} = 9.84$), while J100551.19+125740.6 and J010005.94$-$011059.0 are each within 400 kpc of much smaller galaxies ($\sim 2\%$ of their mass.)  

%

\bibliography{master}

\end{document}